\documentclass{iopconfser}

\usepackage{graphicx}
\usepackage{subcaption}
\usepackage{caption}
\usepackage{amsmath}
\usepackage{graphicx}

\usepackage[colorlinks=true,
            linkcolor=blue,
            citecolor=blue,
            urlcolor=blue]{hyperref}

\usepackage{amsmath,amsfonts}
\usepackage{epsfig}
\usepackage{times}
\usepackage{epsfig}

\usepackage{fancyhdr}
\usepackage{afterpage}
\usepackage{setspace}
\usepackage{float}

\usepackage{amsthm,amssymb}
\usepackage[T1]{fontenc}

\usepackage{xcolor}
\usepackage[linesnumbered, algoruled, vlined]{algorithm2e}

\begin{document}

\title{High-lift Wing Separation Control via Bayesian Optimization and Deep Reinforcement Learning}

\author{R. Montalà$^{1}$, B. Font$^{2}$, O. Lehmkuhl$^{3}$, R. Vinuesa$^{4}$ and I. Rodriguez$^{1}$}

\affil{$^1$TUAREG, Universitat Politècnica de Catalunya (UPC), Terrassa, Spain}
\affil{$^2$Mechanical Engineering, Delft University of Technology (TU Delft), Delft, Netherlands}
\affil{$^3$CASE, Barcelona Supercomputing Center (BSC), Barcelona, Spain}
\affil{$^4$Department of Aeroespace Enigneering, University of Michigan, Ann Arbor, USA}

\email{ricard.montala@upc.edu}

\begin{abstract}
This study investigates active flow control (AFC) of a 30P30N high-lift wing at a Reynolds number $Re_c = 450{,}000$ and angle of attack $\alpha = 23^\circ$ using wall-resolved large-eddy simulations (LES). Two optimization strategies are explored: open-loop Bayesian optimization (BO) and closed-loop deep reinforcement learning (DRL), both targeting the mitigation of stall and the improvement of aerodynamic efficiency via synthetic jets on the slat, main, and flap elements. The uncontrolled configuration was validated against literature data, confirming the reliability of the LES setup. The BO framework successfully identified steady jet velocities that increased efficiency by +10.9\% through a -9.7\% drag reduction while maintaining lift. In contrast, the DRL agent, despite leveraging instantaneous flow information from distributed sensors, achieved only minor improvements in lift and drag, with negligible efficiency gain. Training analysis indicated that the penalty-dominated reward constrained exploration. These results highlight the need for carefully designed rewards and computational acceleration strategies in DRL-based flow control at high Reynolds numbers.
\end{abstract}

\section{Introduction}

High-lift wing configurations are essential for modern transport aircraft during take-off and landing, when multi-element devices such as slats and flaps are deployed to generate lift at low airspeeds. Their deployment, however, introduces complex flow phenomena, including separation in slat and cove regions and wake interactions between wing elements. Aircraft performance metrics such as runway distance, climb rate, payload capacity, and noise emissions are therefore strongly influenced by high-lift aerodynamics \cite{Thibert1995}, making accurate understanding and control of these flows critical for efficient and reliable aircraft design.

The 30P30N high-lift configuration, originally developed by McDonnell Douglas, is a widely used benchmark for the aerodynamics and aeroacoustics of multi-element wings. Introduced in the NASA high-lift prediction workshop \cite{Klausmeyer1997}, early studies combined experiments and RANS simulations to assess predictive capabilities. The configuration later became a reference for aeroacoustic investigations within the AIAA BANC-III workshop \cite{Choudhari2015}, focusing on tonal noise generated by shear-layer impingement in the slat cove at the Reynolds number $Re_c = 1.7\times10^6$ and angle of attack $\alpha = 5.5^\circ$ \cite{Pascioni2014, Ashton2016}. The Reynolds number is defined as $Re_c=U_\infty c/\nu$, with $\nu$ being the fluid kinematic viscosity, $U_\infty$ the freestream velocity, and $c$ the chord of the wing. For high-lift wings, $c$ refers to the nested chord.

More recently, the 30P30N has been used to evaluate advanced numerical methods, including high-order schemes \cite{Gao2020}, mesh refinement techniques \cite{Ueno2019}, and hybrid RANS/LES approaches \cite{Shur2023}. Recent high-fidelity LES studies have further shown that, at high angles of attack, stall originates from shear-layer separation between the main and flap elements \cite{Montala2024, Montala2025}, providing a detailed baseline for flow-control studies.

To mitigate stall, active flow control (AFC) has emerged as a promising technique to manipulate wing flows and enhance aerodynamic performance. Among various AFC methods, synthetic jets, characterized by zero net mass flux, have received significant attention due to their effectiveness in controlling flow separation. Numerical studies have shown that synthetic jets can delay stall and improve efficiency at high angles of attack by mitigating large-scale separation \cite{You2008, Rodriguez2020}. Their applicability has been extended from single-element airfoils to high-lift wing configurations, both experimentally \cite{Melton2006} and numerically \cite{Shmilovich2009}. More recently, Lehmkuhl et al. \cite{Lehmkuhl2020} performed wall-modeled LES on a complete aircraft (JSM model) under stall conditions, demonstrating the feasibility of AFC at full scale.

Nevertheless, selecting optimal AFC parameters remains a significant challenge due to the complex, nonlinear nature of flow dynamics. In recent years, machine learning (ML) has attracted increasing attention, and its combination with AFC has opened new possibilities. In particular, deep reinforcement learning (DRL) has shown potential as a methodology capable of discovering complex actuation strategies to enhance aerodynamic performance. Rabault et al. \cite{Rabault2019} pioneered this approach, achieving drag reduction on a two-dimensional cylinder at $Re_D = 100$ (based on the cylinder diameter $D$). Building on this, Rabault and Kuhnle \cite{Rabault_Kuhnle2019} introduced multi-environment training strategies to improve learning efficiency and scalability to higher Reynolds numbers. More recently, Suárez et al. \cite{Suarez2025a, Suarez2025b} extended DRL to three-dimensional cylinders, achieving significant drag reduction at $Re_D = 3,900$. Beyond bluff body control, DRL has also been applied to reduce skin friction in wall-bounded turbulence \cite{Guastoni2023} and to three-dimensional Rayleigh-Bénard flows \cite{Vasanth2024}, the former representing the first successful DRL study to control turbulence. However, applications to airfoils and wings remain limited, mostly focusing on two-dimensional geometries at low Reynolds numbers \cite{Wang2022, Garcia2025}. More recently, our group has demonstrated the potential of combining DRL with a GPU-accelerated CFD solver \cite{Font2025}. This approach achieved improved performance for a NACA 0012 wing at $Re_c = 1{,}000$ \cite{montala2025_drl_laminar} and an SD7003 wing at $Re_c = 60{,}000$ \cite{montala2025_drl_turb}, highlighting the promise of this methodology for more complex configurations.

Another data-driven methodology to tackle optimization problems is Bayesian optimization (BO). This approach has proven robust in fluid mechanics. Morita et al. \cite{Morita2022} applied BO to a range of shape optimization problems, Mahfoze et al. \cite{Mahfoze2019} used it to reduce skin-friction drag in a turbulent boundary layer via optimum blowing amplitude, Han et al. \cite{Han2023} reduced the drag coefficient of a cylinder at $Re_D=200$ by selecting appropriate steady tangential velocities along the surface, and Li et al. \cite{Li2024} enhanced the jet mixing by optimizing multiple actuation inputs in both the bulk flow and nozzle periphery. Although less advanced than DRL a priori, as it typically relies on fixed control parameters rather than producing a closed-loop control strategy, BO remains an efficient and reliable tool for identifying high-performance configurations.

The present study focuses on the 30P30N geometry at $\alpha = 23^\circ$ and $Re_c=450{,}000$, conditions where stall separation dominates the flow. The objective is to apply an AFC strategy that mitigates this separation and thus improves the wing performance. Two optimization approaches are employed: BO, which systematically explores the actuation parameter space to identify optimal configurations, and DRL, which enables the autonomous discovery of complex, time-dependent control strategies that adapt to the evolving flow dynamics.

\section{Methodology}

In this study, wall-resolved LES are performed for the flow over the 30P30N high-lift wing at a Reynolds number of $Re_c = 450{,}000$ and an angle of attack of $\alpha = 23^\circ$. The objective is to evaluate the effectiveness of synthetic jets in mitigating stall conditions. Two optimization strategies are employed to determine the optimal jet mass flow rate: (i) a classical BO approach, and (ii) a more novel method based on DRL. The Reynolds number has been slightly reduced compared to values reported in the literature to decrease the computational cost of the simulations without altering the flow physics.

\subsection{CFD solver} \label{sec:CFDsetup}
The CFD simulations are performed by solving the filtered incompressible Navier-Stokes equations:

\begin{eqnarray}
\frac{\partial \bar{u_i}}{\partial x_{i}} = 0 \; ,\\
\frac{\partial \bar{u_i}}{\partial t} + \frac{\partial}{\partial x_{j}}(\bar{u}_{i} \bar{u}_{j}) - \nu \frac{\partial^2 \bar{u}_{i}}{\partial{x_{j}} \partial{x_{j}}} + \frac{1}{\rho} \frac{\partial \bar{p}}{\partial x_{i}} = - \frac{\partial \tau_{ij}}{\partial x_{j}} \; ,
\label{eq:LES}
\end{eqnarray}

\noindent where $\overline{u}_i$ ($i=1$, $2$, $3$) and $\overline{p}$ denote the filtered velocity and pressure fields, respectively, $\rho$ is the fluid density, and $\tau_{ij}$ is the subgrid-scale (SGS) stress tensor. The latter represents the effect of unresolved turbulent motions and must therefore be modeled. Its deviatoric part is here modeled through a subgrid-scale viscosity $\nu_\mathrm{SGS}$ as

\begin{equation}
\tau_{ij} - \frac{1}{3}\tau_{kk}\delta_{ij} = -2 \; \nu_\mathrm{SGS} \; \overline{S}_{ij} \; ,
\end{equation}

\noindent where $\overline{S}_{ij} = 1/2 (\partial \overline{u}_i/\partial x_j + \partial \overline{u}_j/\partial x_i)$ is the rate-of-strain tensor of the resolved velocity field, and $\delta_{ij}$ is the Kronecker delta. The subgrid-scale viscosity is here evaluated using the Vreman model \cite{Vreman2004}. The simulations are carried out using SOD2D \cite{Gasparino2024}, an in-house GPU-enabled CFD solver developed at the Barcelona Supercomputing Center, which is based on the spectral element method (SEM).

The three-dimensional computational domain extends $L_x/c = 20$, $L_y/c=20$, and $L_z/c=0.1$ in the streamwise, cross-stream, and spanwise directions, respectively. The 30P30N wing is positioned approximately at the center of the domain and spans the entire width. The domain size is selected based on previous studies \cite{Montala2024,Montala2025}. A uniform freestream velocity $U_\infty$ is imposed at the inlet, with its $x$- and $y$-components determined by the angle of attack $\alpha$. At the outlet, zero-gradient boundary conditions are applied to the velocity field, while the pressure is fixed. A no-slip condition is enforced on all solid surfaces, whereas Dirichlet conditions are used at the jet locations to prescribe jet velocities when AFC is activated.

One synthetic jet is positioned on each wing element. These are located at $x_{\mathrm{jet,slat}}/c=-0.050$, $x_{\mathrm{jet,main}}/c=0.725$, and $x_{\mathrm{jet,flap}}/c=1.0$. These positions are selected to provide the optimization algorithms sufficient flexibility to influence the flow over all elements. The slat and main element jets are positioned to interact with their respective wakes, whereas the flap jet targets the upper-surface recirculation region. The jet angles are set to $\phi_\mathrm{jet} = 90^\circ$ for suction and $\phi_\mathrm{jet} = 20^\circ$ for blowing, measured with respect to the wall-tangential direction. The streamwise jet width is $h_\mathrm{jet}/c = 0.01$, and each jet spans the full domain width along the spanwise direction.

In this study, the velocities of these synthetic jets are determined through optimization. Specifically, BO is used to select a fixed, steady jet velocity, while DRL provides closed-loop control, allowing the agent to adjust the jet velocity instantaneously based on the evolving flow state. In both cases, the optimization determines the appropriate velocity for each jet, but DRL offers dynamic control, whereas BO yields a single optimal steady velocity.

\subsection{Bayesian optimization}\label{sec:BOsetup}

For the BO process, fixed jet velocities are applied to the slat and main jets, while the velocity of the flap jet is constrained to ensure mass conservation, being set to produce an opposite mass flow rate equal to the sum of the other two jets:

\begin{eqnarray}
\label{eq:flap_jet}
U_{\mathrm{jet,\,flap}} = - \left( U_{\mathrm{jet,\,slat}} + U_{\mathrm{jet,\,main}} \right).
\end{eqnarray}

Therefore, two control parameters are optimized during the BO process: the slat and main jet velocities. In each simulation, fixed values of jet velocities are prescribed, where the exploration range is set to $U_{\mathrm{jet,\,slat}}/U_\infty \in [-1.22, 1.22]$ and $U_{\mathrm{jet,\,main}}/U_\infty \in [-1.22, 1.22]$ to prevent impractical actuations in which the jet velocities would significantly exceed the freestream velocity. Additionally, this maximum magnitude is based on the results by Shmilovich and Yadlin \cite{Shmilovich2009}, who used a momentum coefficient of $C_\mu= (h_\mathrm{jet} / c)(U_\mathrm{jet, max}^2 / U_\infty^2) =0.015$ in their AFC numerical simulations.

The objective (or target) function to be optimized in the BO framework is defined as

\begin{equation}
\label{eq:target}
r = \frac{E-E_\mathrm{base}}{E_\mathrm{base}} - W_{C_l} \;P_{C_l} - W_{C_d} \; P_{C_d} \; ,
\end{equation}

\noindent where

\begin{equation}
\label{eq:target_penaltyCl}
P_{C_l} =
\begin{cases}
  \dfrac{C_{l, \mathrm{base}} - C_l}{C_{l, \mathrm{base}}}, & \text{if } C_l < C_{l, \mathrm{base}}, \\
  0, & \text{otherwise.}
\end{cases}
\end{equation}

\noindent and

\begin{equation}
\label{eq:target_penaltyCd}
P_{C_d} =
\begin{cases}
  \dfrac{C_d - C_{d, \mathrm{base}}}{C_{d, \mathrm{base}}}, & \text{if } C_d > C_{d, \mathrm{base}}, \\
  0, & \text{otherwise.}
\end{cases}
\end{equation}

The objective $r$ aims to maximize the aerodynamic efficiency $E=C_l/C_d$ relative to the non-actuated baseline case $E_\mathrm{base}$. However, improvements in $E$ may lead to undesirable trade-offs, such as reduced lift or increased drag. To penalize such effects, penalty terms $P_{C_l}$ and $P_{C_d}$ are introduced whenever the lift coefficient decreases, or the drag coefficient increases relative to the uncontrolled scenario. These penalties are expressed as relative deviations and are weighted by $W_{C_l}$ and $W_{C_d}$ in the objective function. In this work, both weights are set to $W_{C_l} = 100$ and $W_{C_d} = 100$ to impose strong constraints, ensuring that neither lift reduction nor drag increase occurs in the optimized configuration.

The relationship between the control parameters $\mathbf{v} = [U_{\mathrm{jet,\,slat}}, U_{\mathrm{jet,\,main}}]$ and the objective $r$ is modeled using Gaussian process (GP) surrogate models. A GP is a probabilistic model used to approximate an unknown function, here representing aerodynamic quantities such as $C_l = f_l(\mathbf{v})$, $C_d = f_d(\mathbf{v})$, and $E= f_E(\mathbf{v})$, which are combined to construct the objective function $r$ defined above. Then, at any new control input $\mathbf{v}^* = [U_{\mathrm{jet,\,slat}}, U_{\mathrm{jet,\,main}}]$, the GP provides a probabilistic prediction of the form

\begin{equation}
f(\mathbf{v}^*) \sim
\mathcal{N}\big(\mu(\mathbf{v}^*), \sigma^2(\mathbf{v}^*)\big),
\end{equation}

\noindent for each aerodynamic quantify, where $\mu(\mathbf{v}^*)$ is the predicted mean (surrogate estimate) and $\sigma(\mathbf{v}^*)$ is the standard deviation, representing the model uncertainty. These quantities are computed as

\begin{equation}
\mu(\mathbf{v}^*) =
\boldsymbol{\kappa}^T (K + \sigma_n^2 I)^{-1} \mathbf{y},
\qquad
\sigma^2(\mathbf{v}^*) =
k(\mathbf{v}^*, \mathbf{v}^*) -
\boldsymbol{\kappa}^T (K + \sigma_n^2 I)^{-1} \boldsymbol{\kappa}.
\end{equation}

Here $\mathbf{y} = [y_1, \dots, y_N]^T$ are the outputs of the previous $N$ CFD simulations, corresponding to the training set $\{(\mathbf{v}_i, y_i)\}_{i=1}^{N}$. The covariance matrix $K \in \mathbb{R}^{N \times N}$ contains the covariances between all training points $K_{ij} = k(\mathbf{v}_i, \mathbf{v}_j)$, where $k(\mathbf{v}_i, \mathbf{v}_j)$ is the kernel function. In this work, a
radial basis function (RBF) kernel is employed,

\begin{equation}
k(\mathbf{v}_i, \mathbf{v}_j) =
\exp\!\left(
-\frac{\|\mathbf{v}_i - \mathbf{v}_j\|^2}{2 l^2}
\right),
\end{equation}

\noindent with $\|\mathbf{v}_i - \mathbf{v}_j\|$ denoting the Euclidean distance between the two input vectors and $l$ the kernel length-scale controlling the decay of correlation. The vector $\boldsymbol{\kappa} =
\big[k(\mathbf{v}_1, \mathbf{v}^*), \dots, k(\mathbf{v}_N, \mathbf{v}^*)\big]^T$ represents the covariance between the new input $\mathbf{v}^*$ and each training point $\mathbf{v}_i$, while $k(\mathbf{v}^*, \mathbf{v}^*)$ denotes the self-covariance of the new input. The noise variance $\sigma_n^2$ accounts for numerical uncertainty in the CFD outputs, and both $\sigma_n^2$ and $l$ are fitted during GP training. Each newly sampled configuration $\mathbf{v}^*$ updates the GP model at each BO iteration. The predicted mean guides the optimization toward promising regions of the control space, while the associated uncertainty promotes exploration of under-sampled regions, ensuring efficient convergence toward the optimal jet configuration.

\subsection{DRL optimization}\label{sec:DRLsetup}

\begin{figure}[h]
  \centering
    \includegraphics[width=\linewidth]{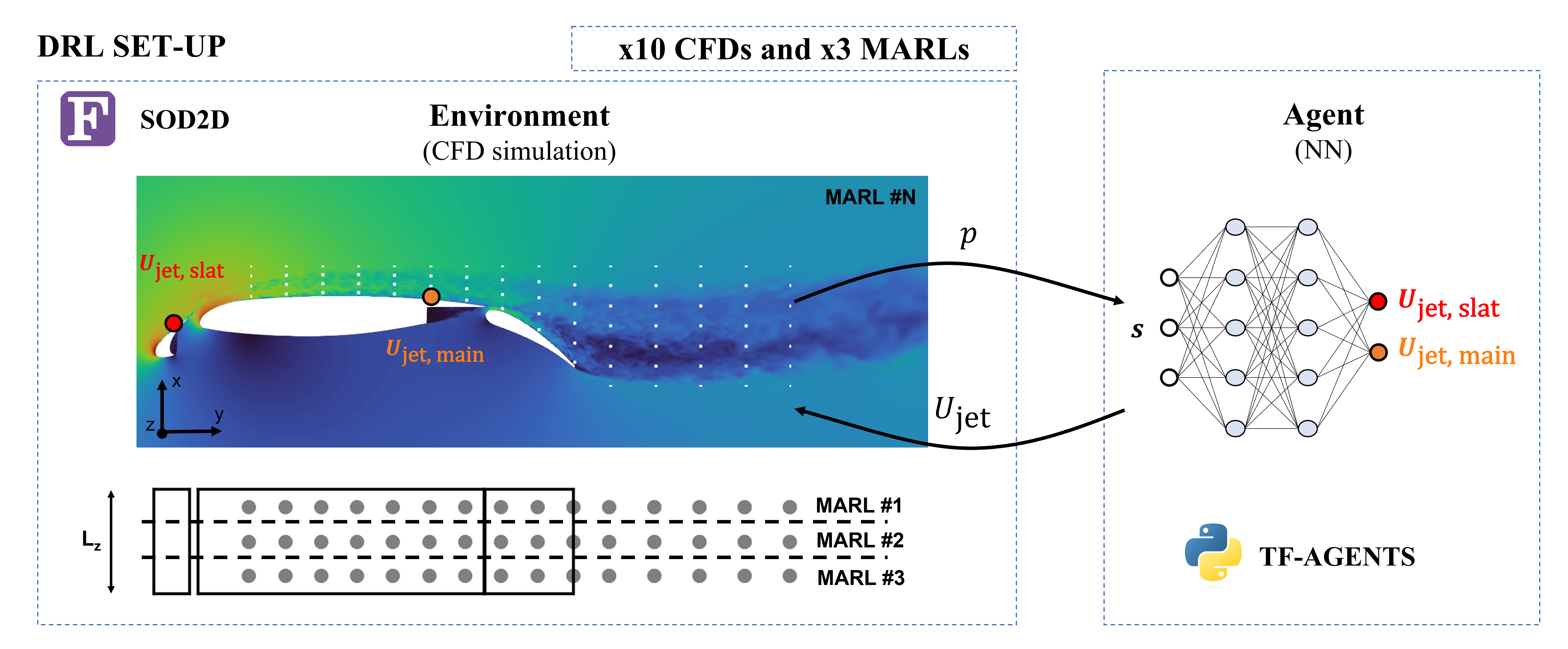}
  \caption{CFD-DRL set-up}
  \label{fig:DRLsetup}
\end{figure}

The DRL framework, illustrated in Figure \ref{fig:DRLsetup}, consists of two main components: the CFD environment, which was presented in Section \ref{sec:CFDsetup}, and the DRL agent. The latter is implemented as a neural network (NN) and maps flow states to control actions. In this setup, the NN receives pressure measurements from sensors within the CFD environment and outputs the slat and main jet velocities ($U_\mathrm{jet,slat}$ and  $U_\mathrm{jet,main}$) to be applied back to the system. The flap jet is constrained to operate at an opposing mass flow rate relative to the other jets, as defined in Eq. \ref{eq:flap_jet}. The agent is implemented using the TF-Agents Python library \cite{Guadarrama2018} and communicates with the Fortran-based CFD solver via a Redis in-memory database managed by SmartSim \cite{Partee2022}, enabling efficient coupling and online training of the agent.

To enable three-dimensional actuation without excessive computational cost, a multi-agent reinforcement learning (MARL) framework is employed \cite{Belus2019, Suarez2025a, Suarez2025b, Font2025}. The computational domain is then divided into multiple pseudo-environments, each containing its own set of jets (slat, main, and flap), while all pseudo-environments share a single agent. Thus, based on local flow states, the agent outputs distinct jet velocities for each pseudo-environment, allowing spatially distributed control with manageable training complexity. In this study, three pseudo-environments are used ($n_\text{MARL} = 3$), each spanning $L_{z,\mathrm{MARL}}/c = 0.1$, so that the DRL domain is three times larger ($L_z/c = 0.3$) than in the baseline configuration.

To accelerate data collection and reduce wall-clock time, ten independent CFD simulations run concurrently ($n_\text{CFD} = 10$). At the end of each episode, a total of $3 \times 10 = 30$ state-action-reward trajectories are collected for policy updates. Each episode lasts $T_\text{eps} \, U_\infty/c = 8.57$, corresponding to twelve vortex-shedding cycles in the uncontrolled case. A total of 120 actions are applied per pseudo-environment, yielding an action duration of $T_\text{act} = T_\text{eps}/120$.

Each pseudo-environment contains 78 sensors in the $z$-midplane (see Figure \ref{fig:DRLsetup}). To provide three-dimensional flow information, the NN input includes sensor slices from adjacent pseudo-environments, producing a total state vector of size $78 \times 3 = 234$.

The proximal policy optimization (PPO) algorithm \cite{Schulman2017} is used to refine the agent's policy. Both the actor and critic networks are implemented as multi-layer perceptrons (MLPs) with two hidden layers of 512 neurons each. The actor maps observed states to actions, while the critic estimates the advantage function, quantifying the relative benefit of an action compared to the current policy. The critic is trained using trajectory data at the end of each episode, and the actor is subsequently updated based on these advantage estimates, with PPO's clipped objective ensuring stable policy updates.

The reward function for DRL is based on the same target quantity used in the BO optimization, as defined in Eq. \ref{eq:target}. Within the MARL framework, each pseudo-environment computes a local reward $r_i$, which is then combined with the rewards from the other pseudo-environments to form the final training reward $R_i$:

\begin{equation} \label{eq:Reward_global}
	R_i = \gamma \, r_i + \frac{1-\gamma}{n_\text{MARL}} \sum_{j=1}^{n_\text{MARL}} r_j,
\end{equation}

\noindent where the weighting factor $\gamma$ balances the relative importance of local versus averaged global performance. In this work, this is set to $\gamma=0.8$.

\section{Results}
This section first validates the non-actuated scenario to ensure the reliability of the subsequent BO and DRL results (Section \ref{sec:baseline_results}). It also establishes a reference case to facilitate comparison and quantify the performance improvements achieved by the optimization processes. Following this, the AFC results obtained using the BO and DRL methodologies are analyzed in Sections \ref{sec:BO_results} and \ref{sec:DRL_results}, respectively.

\subsection{Baseline validation} \label{sec:baseline_results}
In this section, the baseline case without AFC is validated against numerical results available in the literature. Since no studies at the same Reynolds number are present in the literature, the results from Montalà et al. \cite{Montala2024,Montala2025} obtained at a slightly higher Reynolds number ($Re_c = 750{,}000$) and the same angle of attack ($\alpha = 23^\circ$) are used for comparison. Although minor discrepancies may arise due to the difference in Reynolds number, the overall flow physics are expected to be similar, and thus only small variations in the results are anticipated.

In this work, two mesh refinement levels are employed. Since both optimization strategies (BO and DRL) require running multiple CFD simulations with varying parameters, a coarse mesh is used during model training, while the final evaluations are performed on the fine mesh. This approach significantly reduces the overall computational cost of both methodologies. This mesh refinement is achieved through \textit{p}-refinement, i.e., by increasing the polynomial order of the high-order basis functions used in SEM. For the coarse mesh, a polynomial order of $p = 2$ is used, while for the fine mesh this is increased to $p = 4$. Consequently, the topology and geometry of the elements remain identical between the two meshes, and only the number of grid points within each element increases from the coarse to the fine mesh.

The mesh characteristics are summarized in Table \ref{tab:mesh_info}. The table also includes the dimensionless wall distances of the mesh. These quantities are evaluated in the fully turbulent regions of the flow along the suction sides of the elements, and both the average and maximum values are reported. Although the coarse mesh slightly exceeds the recommended limits for wall-resolved LES, particularly the $\Delta y^+$ distances, the fine mesh lies well within the accepted ranges \cite{Piomelli1996}, ensuring that $\Delta x^+ < 150$, $\Delta y^+ < 2$, and $\Delta z^+ < 40$ across the entire wall surface where turbulence is present.

\begin{table}[h]
\begin{center}
\begin{tabular}{lccccccccc}
\hline
 & $N_{xy}$ & $N_{z}$ & $N_{xyz}$ & $\Delta x^{+}_\mathrm{max}$ & $\Delta y^{+}_\mathrm{max}$ & $\Delta z^{+}_\mathrm{max}$ & $\Delta x^{+}_\mathrm{avg}$ & $\Delta y^{+}_\mathrm{avg}$ & $\Delta z^{+}_\mathrm{avg}$  \\ \hline
Coarse Mesh & $2.10 \times 10^{5}$ & 49 & $10$ M & 125 & 3.5 & 70 & 72 & 2.2 & 48\\
Fine Mesh & $8.38 \times 10^{5}$ & 97 & $81$ M & 85 & 1.5 & 40 & 43 & 0.9 & 28 \\ \hline
\end{tabular}
\caption{Mesh information for the two mesh refinements considered, including the number of grid points in the $x$--$y$ plane $N_{xy}$, along the $z$-direction $N_z$, and the total number of points $N_{xyz}$ (in millions). Also, the maximum and average wall-tangential $\Delta x^+$, wall-normal $\Delta y^+$, and spanwise $\Delta z^+$ distances are reported.}
\label{tab:mesh_info}
\end{center}
\end{table}

\begin{figure}[h]
  \centering
  \begin{subfigure}[b]{0.49\textwidth}
    \centering
    \includegraphics[width=\linewidth]{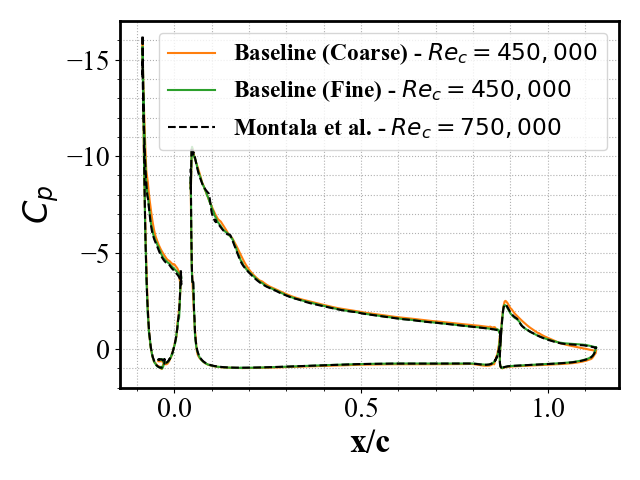}
    \caption{}
    \label{fig:Baseline_coeff_distribution_a}
  \end{subfigure}
  \begin{subfigure}[b]{0.49\textwidth}
    \centering
    \includegraphics[width=\linewidth]{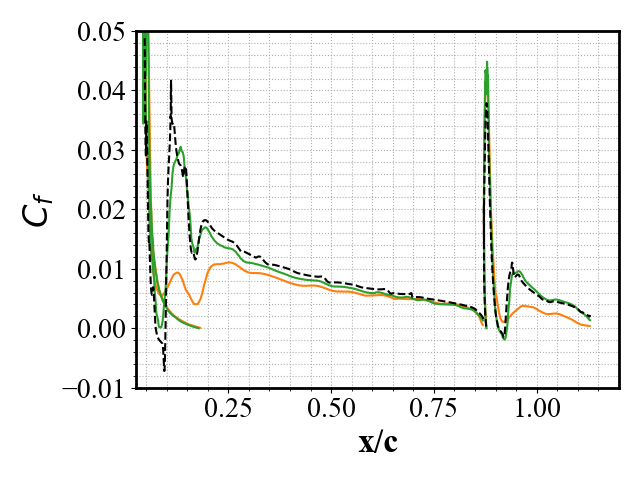}
    \caption{}
    \label{fig:Baseline_coeff_distribution_b}
  \end{subfigure}
  \caption{Pressure coefficient $C_p$ (left) and skin friction coefficient $C_f$ (right) distributions for the non-actuated case over the walls of the wing. The skin friction coefficient is only shown on the suction side of the main and flap elements.}
  \label{fig:Baseline_coeff_distribution}
\end{figure}

To assess differences with the literature, Figure \ref{fig:Baseline_coeff_distribution} shows the distributions of the pressure coefficient $C_p$ and skin friction coefficient $C_f$ over the surfaces of each element. From this figure, it is evident that the fine mesh exhibits better agreement with the reference results than the coarse mesh, even though some discrepancies are anticipated due to the differences in the Reynolds number. Differences are especially pronounced for the skin friction coefficient, which is more sensitive to Reynolds number and near-wall resolution. The coarse mesh exhibits larger deviations due to its higher $\Delta y^+$, although both meshes correctly capture overall trends and transition locations. Since the flow is dominated by pressure forces at this high angle of attack, discrepancies in skin friction have only a minor effect on the overall aerodynamic coefficients.

\begin{table}[h]
\begin{center}
\begin{tabular}{lccccc}
\hline
CASE           & $C_l$     & $C^\prime_{l, \mathrm{rms}}$ & $C_d$ & $C^\prime_{d, \mathrm{rms}}$ & $E$ \\ \hline
Baseline Coarse    & 3.9555 & 0.0265 & 0.3627 & 0.0069 & 10.91 \\
Baseline Fine   & 3.9187 & 0.0266 & 0.2448 & 0.0118 &  16.01 \\
Montalà et al. \cite{Montala2024,Montala2025} & 4.0627 & -      & 0.1983 & - & 20.49  \\ \hline
\end{tabular}
\caption{Time-averaged aerodynamic coefficients for the non-actuated case are shown for both the coarse and fine meshes, compared against values reported in the literature. Each column displays the lift coefficient $C_l$, the root-mean-square of the lift coefficient fluctuations $C'_{l,\mathrm{rms}}$, the drag coefficient $C_d$, and the root-mean-square of the drag coefficient fluctuations $C'_{d,\mathrm{rms}}$.}
\label{tab:Baseline_coeff}
\end{center}
\end{table}

The time-averaged aerodynamic coefficients obtained with each mesh are reported in Table \ref{tab:Baseline_coeff}, including the lift $C_l$ and drag $C_d$ coefficients, along with the root-mean-square values of their fluctuations, $C^\prime_{l, \mathrm{rms}}$ and $C^\prime_{d, \mathrm{rms}}$, respectively. From this table, it is evident that, despite the considerably lower resolution of the coarse mesh, it still predicts the lift coefficient reasonably well. Since the lift is mainly dominated by the pressure contribution, the difference between the coarse and fine meshes is less than 1\%. The drag coefficient, however, shows a much larger discrepancy, with a difference of approximately 48\%. This is because the skin friction contribution, which is more sensitive to mesh resolution, is comparatively higher in the drag coefficient, leading to greater differences between the coarse and fine meshes. Additionally, the offset in the pressure coefficient distribution downstream of the suction peaks further contributes to this discrepancy (see Figure \ref{fig:Baseline_coeff_distribution}). When comparing the rms values, a similar behavior is observed: $C'_{l,\mathrm{rms}}$ is similar for both the coarse and fine meshes, whereas the difference in $C'_{d,\mathrm{rms}}$ is considerably larger. Nevertheless, this is not considered critical, as the coarse mesh is used only for training the models and not for their final evaluation.

\begin{figure}[h]
  \centering
  \begin{subfigure}[b]{0.49\textwidth}
    \centering
    \includegraphics[width=\linewidth]{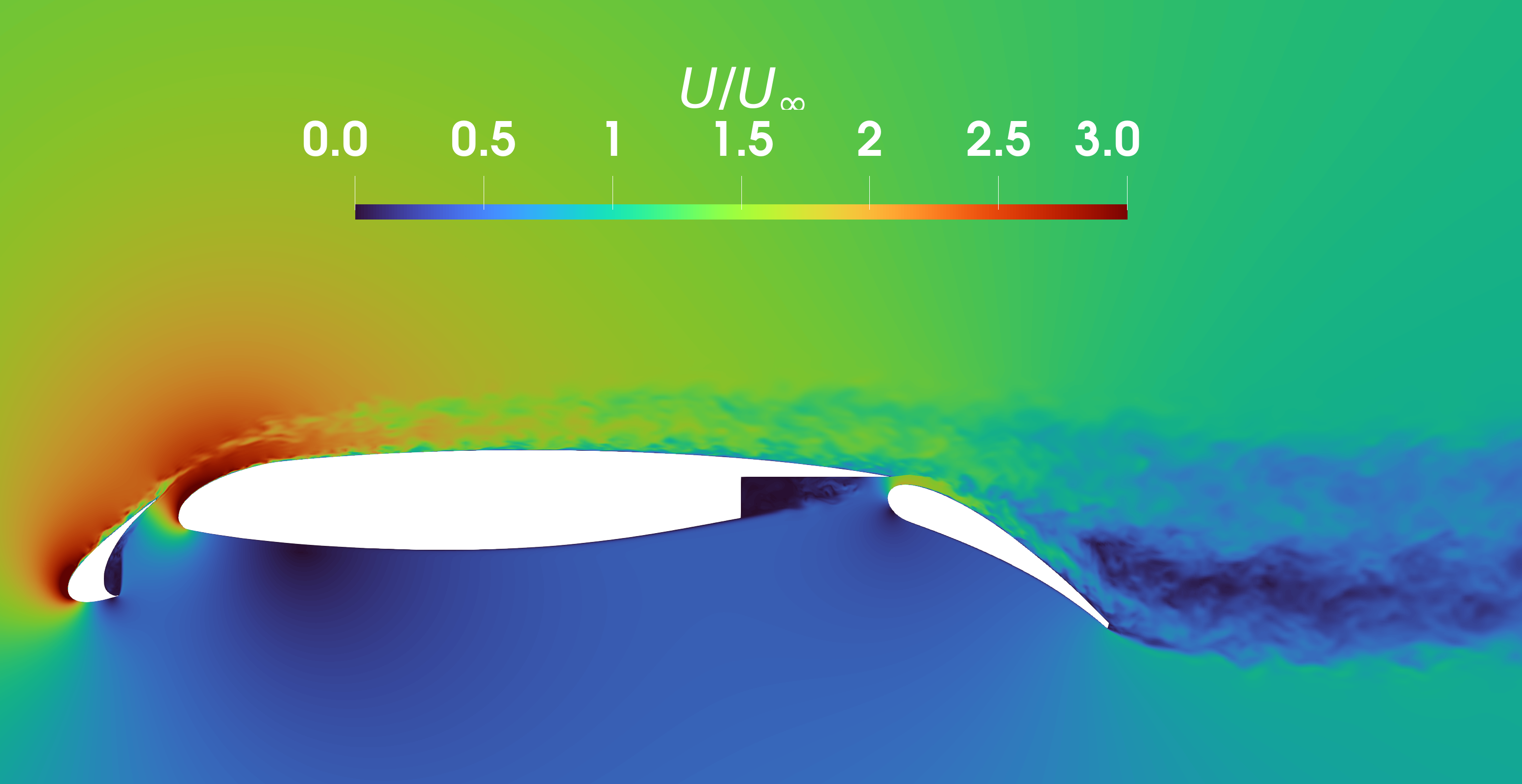}
    \caption{}
    \label{fig:Baseline_flow_a}
  \end{subfigure}
  \begin{subfigure}[b]{0.49\textwidth}
    \centering
    \includegraphics[width=\linewidth]{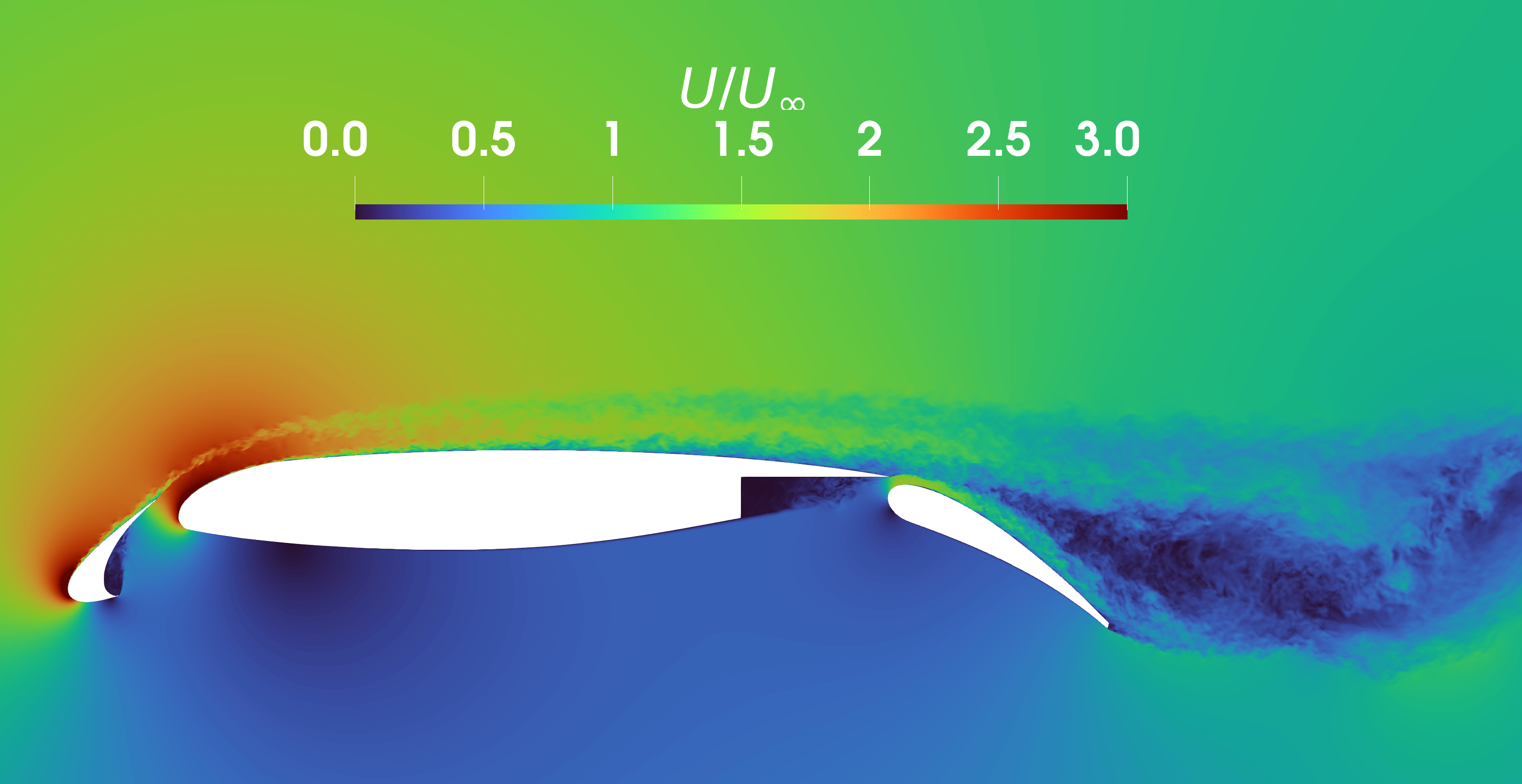}
    \caption{}
    \label{fig:Baseline_flow_b}
  \end{subfigure} 
  \\
    \begin{subfigure}[b]{0.49\textwidth}
    \centering
    \includegraphics[width=\linewidth]{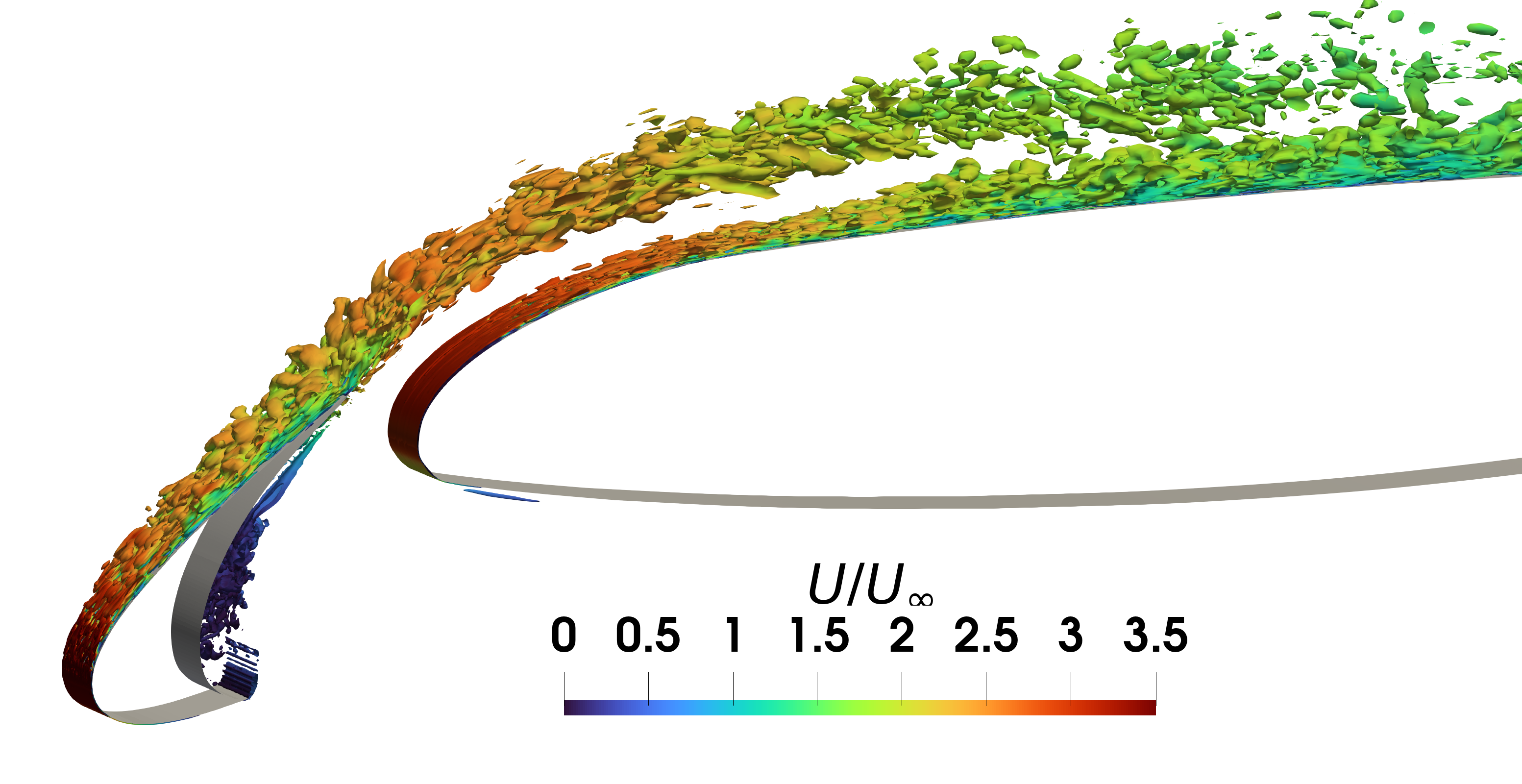}
    \caption{}
    \label{fig:Baseline_flow_c}
  \end{subfigure}
  \begin{subfigure}[b]{0.49\textwidth}
    \centering
    \includegraphics[width=\linewidth]{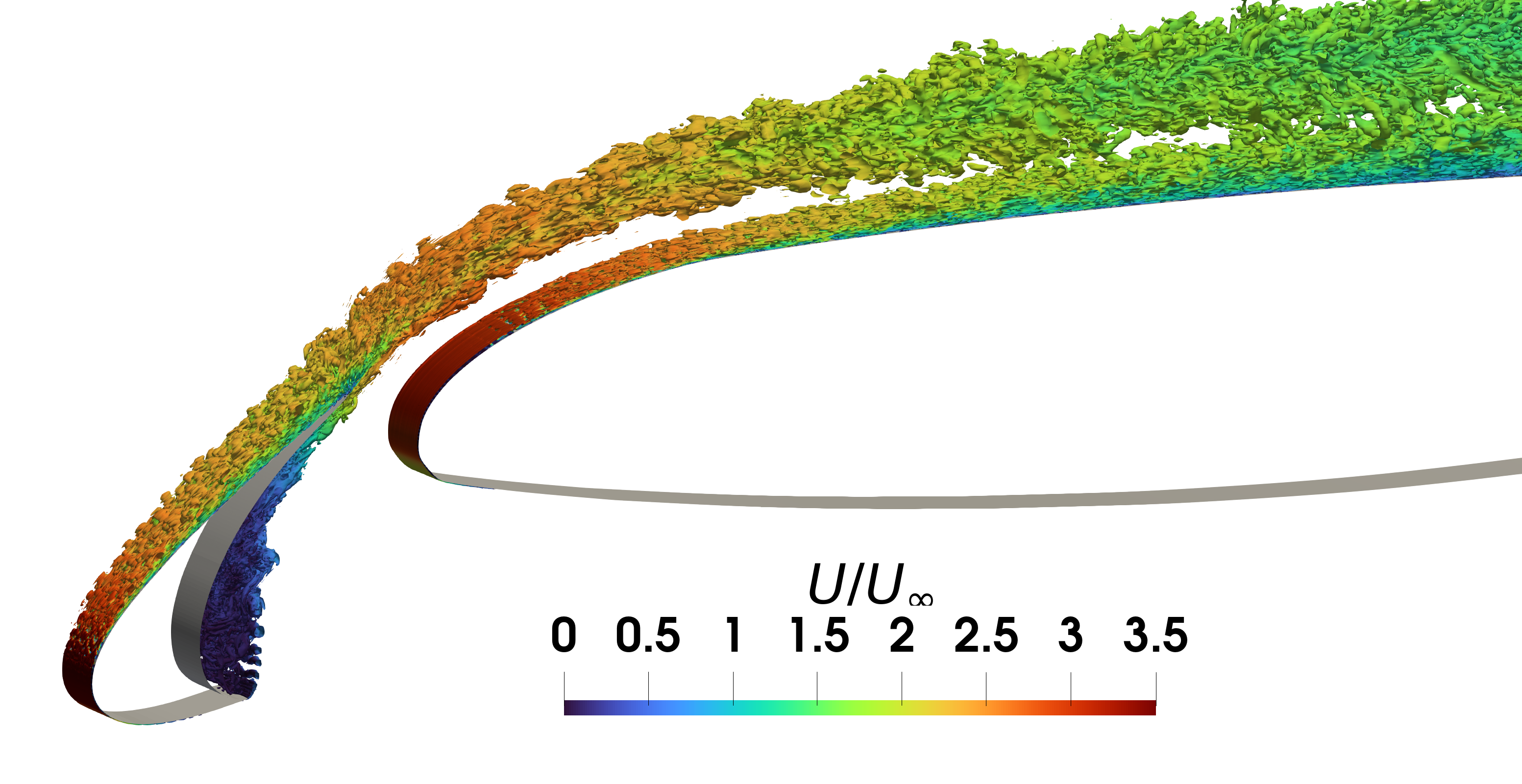}
    \caption{}
    \label{fig:Baseline_flow_d}
  \end{subfigure}
  \caption{Instantaneous fields from the coarse (left panel) and fine meshes (right panel) in the uncontrolled scenario. The top rows show the velocity magnitude field in the mid-$z$ plane, while the bottom rows depict the vortical structures visualized using Q-criterion iso-contours, colored by velocity magnitude.}
  \label{fig:Baseline_flow}
\end{figure}

Figure \ref{fig:Baseline_flow} shows instantaneous flow visualizations. In these images, the higher resolution of the fine mesh is evident, with many more small-scale details captured compared to the coarse mesh. Nevertheless, the same physical features are present in both grids despite the differences in accuracy. The recirculation bubbles in the main and slat cove regions have roughly the same size, the laminar-to-turbulent transition regions occur at the same positions on the leading edges of each element, and similar wakes are observed. All these flow phenomena were thoroughly explained in the work of Montalà et al. \cite{Montala2024,Montala2025}, and the results here are consistent with their observations.

The main difference between the coarse and fine meshes appears in the low-velocity wake region above the flap. In the fine mesh, the wake exhibits a pronounced curvature, which is a signature of vortex shedding. However, this is not captured by the coarse mesh and, although the main statistical quantities are accurately captured by both meshes, as discussed in Table \ref{tab:Baseline_coeff}, the coarse mesh does not fully resolve the instantaneous dynamics. Consequently, the characteristic peak associated with vortex shedding in the Fourier transform of the lift signal is absent in the coarse mesh. Nevertheless, this limitation is not considered important, as the time-averaged coefficients used for training the models are still well represented.

\subsection{Bayesian optimization} \label{sec:BO_results}

Initially, 15 cases were evaluated to construct the GP models, which were iteratively refined as the BO algorithm proposed new sampling points. The optimization proceeded until convergence, defined as the situation in which a newly proposed point was very close to the previous one and the associated uncertainty in the optimization space was nearly eliminated. In total, 4 additional simulations were required beyond the initial samples, resulting in 19 evaluated cases. All simulations were performed using the coarse mesh described in Section \ref{sec:baseline_results}.

\begin{figure}[h]
  \centering
  \begin{subfigure}[b]{0.38\textwidth}
    \centering
    \includegraphics[width=\linewidth]{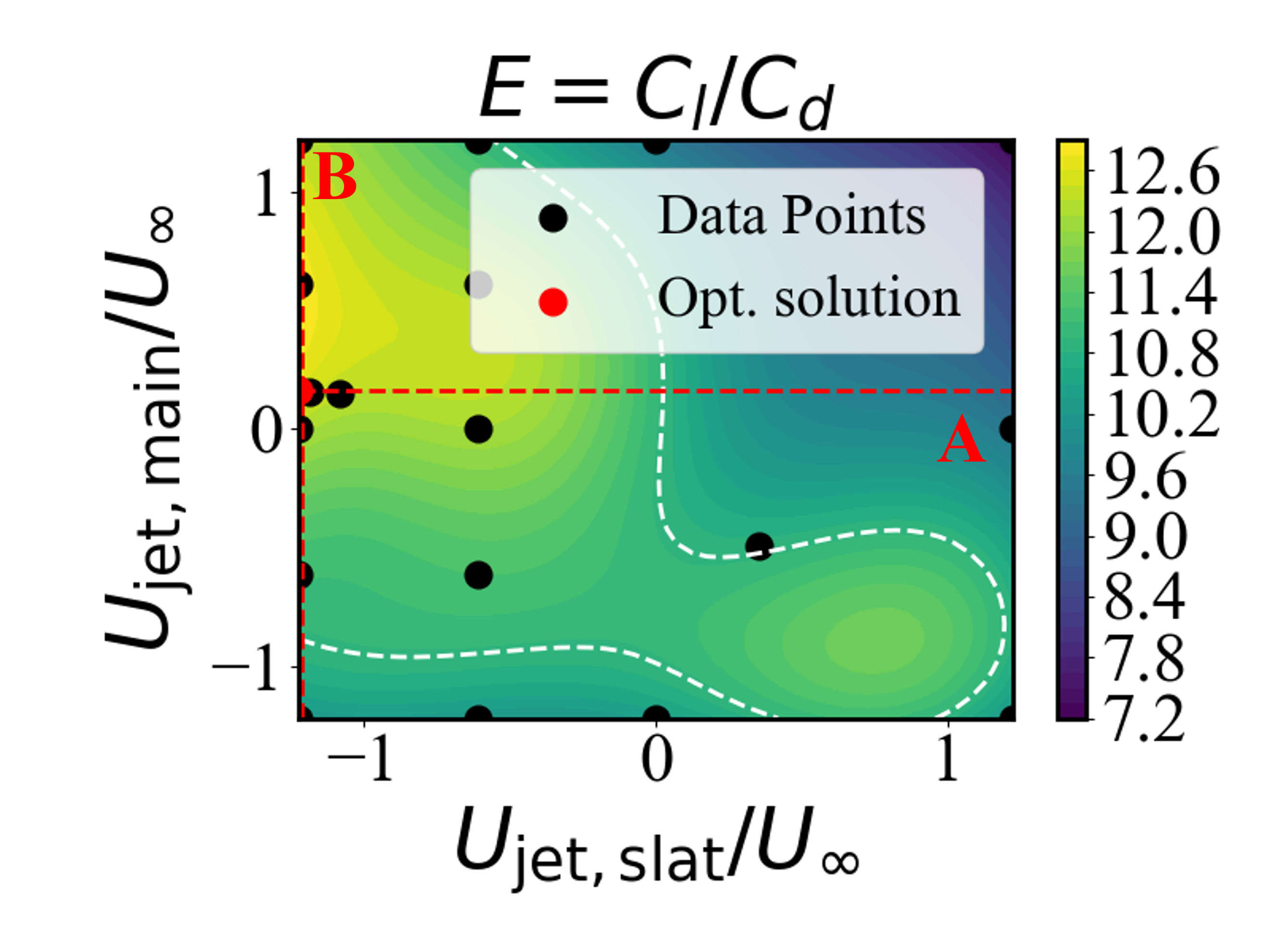}
    \caption{}
    \label{fig:BO_manifold_a}
  \end{subfigure}
  \begin{subfigure}[b]{0.295\textwidth}
    \centering
    \includegraphics[width=\linewidth]{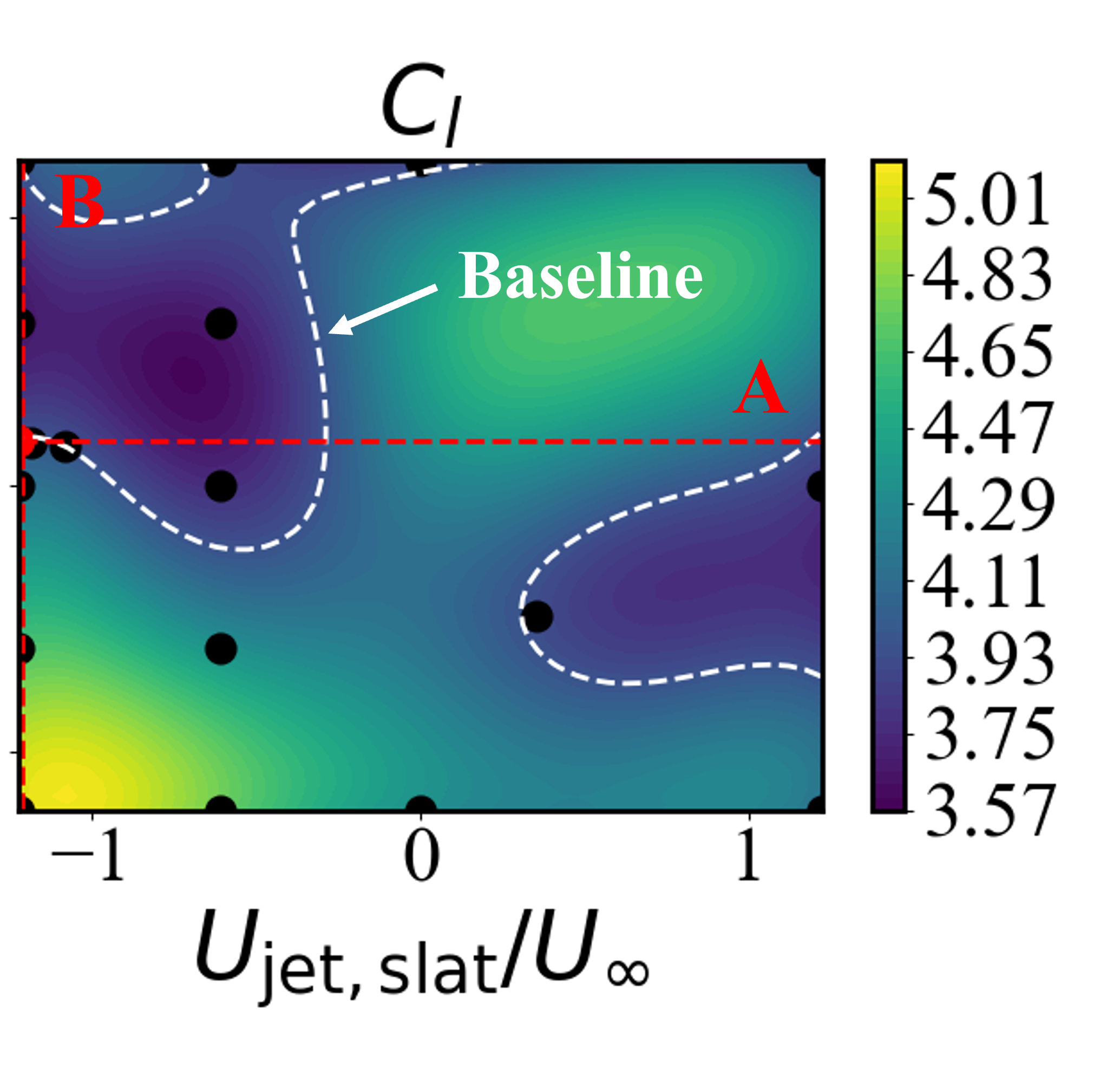}
    \caption{}
    \label{fig:BO_manifold_b}
  \end{subfigure} 
    \begin{subfigure}[b]{0.295\textwidth}
    \centering
    \includegraphics[width=\linewidth]{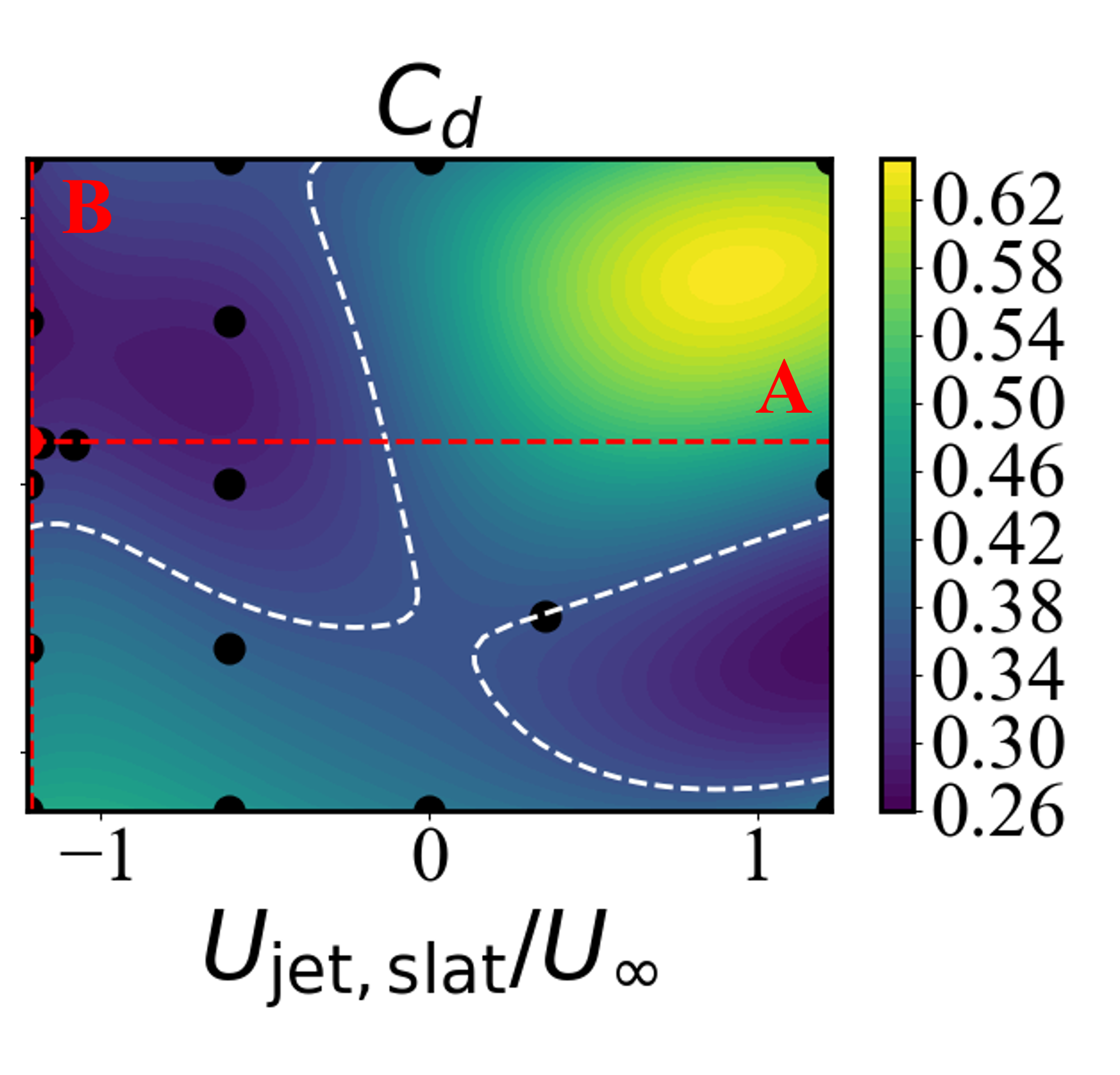}
    \caption{}
    \label{fig:BO_manifold_c}
  \end{subfigure}
  \caption{Efficiency $E$ (left), lift coefficient $C_l$ (middle), and drag coefficient $C_d$ (right) manifolds predicted by the GP models as a function of the slat and main jet velocities.}
  \label{fig:BO_manifold}
\end{figure}

Figure \ref{fig:BO_manifold} shows the aerodynamic efficiency $E$, lift coefficient $C_l$, and drag coefficient $C_d$ isocontours as functions of the slat and main jet velocities ($U_{\mathrm{jet,\,slat}}$ and $U_{\mathrm{jet,\,main}}$) predicted by the GP models. The black points indicate the data used to train the model, while the red point represents the optimum identified by the BO algorithm. White dashed lines are also included. These lines indicate the combinations of slat and flap jet actuation that yield the same aerodynamic coefficients as the non-actuated case (see Table \ref{tab:Baseline_coeff}). These plots confirm that the BO approach successfully increases aerodynamic efficiency $E$ while remaining in the region of the design space where $C_l > C_{l,\mathrm{base}}$ and $C_d < C_{d,\mathrm{base}}$, which is consistent with the constraints imposed in the target function defined in Eq. \ref{eq:target}.

\begin{figure}[h]
  \centering
  \begin{subfigure}[b]{0.8\textwidth}
    \centering
    \includegraphics[width=\linewidth]{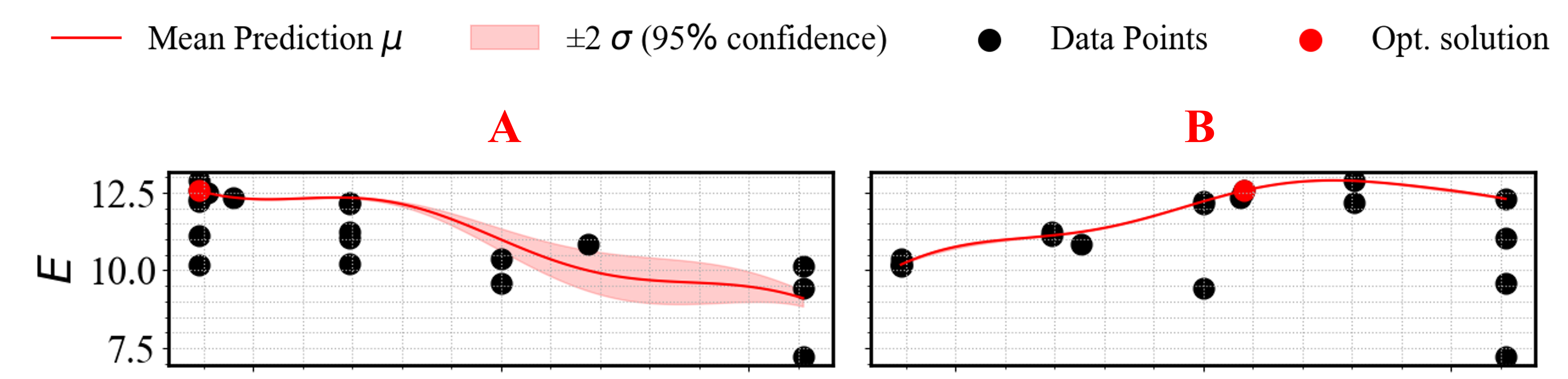}
    \caption{}
    \label{fig:BO_manifold_slice_a}
  \end{subfigure}
  \\
  \begin{subfigure}[b]{0.8\textwidth}
    \centering
    \includegraphics[width=\linewidth]{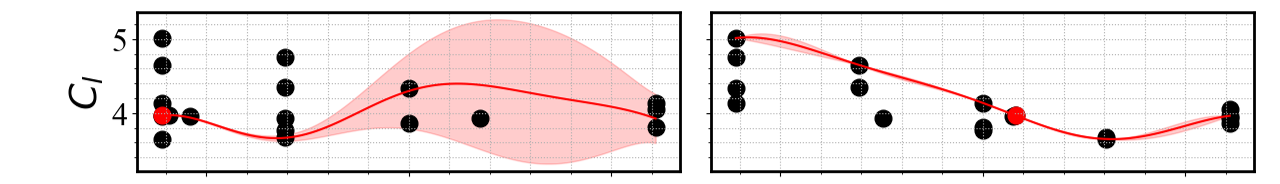}
    \caption{}
    \label{fig:BO_manifold_slice_b}
  \end{subfigure} 
  \\
    \begin{subfigure}[b]{0.8\textwidth}
    \centering
    \includegraphics[width=\linewidth]{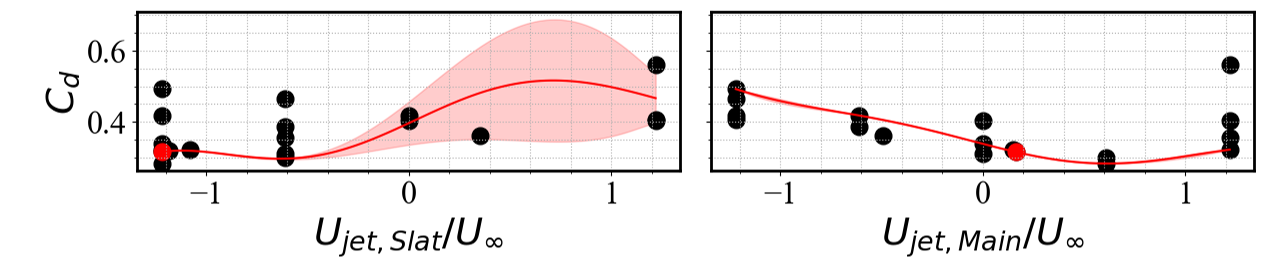}
    \caption{}
    \label{fig:BO_manifold_slice_c}
  \end{subfigure}
  \caption{Slice A (left panel) and B (right panel) of the efficiency $E$ (top), lift coefficient $C_l$ (middle), and drag coefficient $C_d$ (bottom) manifolds predicted by the GP models, corresponding to cross-sections through the optimal point in Figure \ref{fig:BO_manifold}.
}
  \label{fig:BO_manifold_slice}
\end{figure}

To further illustrate the uncertainty across the design space, Figure \ref{fig:BO_manifold_slice} presents two slices (A and B) highlighted in Figure \ref{fig:BO_manifold} as red dashed lines, corresponding to cross-sections through the optimal point along each direction. It is evident that the uncertainty at the optimal point is nearly zero, while regions with fewer data points and less exploration exhibit noticeably higher uncertainty. Moreover, even in these high-uncertainty regions, the aerodynamic efficiency is lower with 95\% confidence. These slices also show that the maximum efficiency is not located exactly at the optimal point; however, the selected optimum satisfies the constraints imposed by the target function. In the region of maximum $E$, the drag coefficient is lower than at the optimal point, but the lift coefficient is also significantly reduced, even below the baseline value, violating the constraint. Therefore, the BO algorithm successfully balances efficiency improvement with the lift and drag constraints.

\begin{table}[h]
\begin{center}
\resizebox{\textwidth}{!}{%
\begin{tabular}{lcccc}
\hline
CASE           & $C_l$     & $C^\prime_{l, \mathrm{rms}}$ & $C_d$ & $E$ \\ \hline
Opt. solution (Coarse) & 3.9654 (+0.25\%) & 0.0299 (+12.83\%) & 0.3157 (-12.97\%) & 12.56 (+15.19\%) \\
Opt. solution (Fine) & 3.9244 (+0.15\%) & 0.0289 (+8.30\%) & 0.2210 (-9.69\%) & 17.76 (+10.89\%) \\ \hline
\end{tabular}
}
\caption{Time-averaged aerodynamic coefficients for the coarse and fine meshes in the BO-optimized AFC scenario. Each column displays the lift coefficient $C_l$, the root-mean-square of the lift coefficient fluctuations $C'_{l,\mathrm{rms}}$, the drag coefficient $C_d$, and the aerodynamic efficiency $E$, along with the relative changes with respect to the baseline scenario for each mesh.}
\label{tab:BO_coeff}
\end{center}
\end{table}

The results for the optimal point are reported in Table \ref{tab:BO_coeff}. Once the optimal point was identified, the results were also evaluated on the fine mesh to ensure the reliability of the optimum. This table therefore presents the results for both the coarse and fine meshes, evaluated at the optimal point shown in Figures \ref{fig:BO_manifold} and \ref{fig:BO_manifold_slice}, i.e. $U_\mathrm{jet,slat}/U_\infty=-1.22$ and $U_\mathrm{jet,main}/U_\infty=0.16$

Despite the expected differences between the coarse and fine meshes, the coarse mesh is able to provide reasonable predictions that translate into an improvement in wing performance on the fine mesh, confirming the reliability of the BO study conducted with the coarse grid. Although the final efficiency achieved on the fine mesh is slightly lower than that on the coarse mesh, compared to the respective baseline scenarios, the results are still promising: the aerodynamic efficiency increases by up to 10.89\% compared with the baseline scenario, while the lift is maintained (+0.15\%) and the drag is substantially reduced (-9.69\%), showing trends similar to those observed on the coarse mesh. Therefore, the improvement in aerodynamic efficiency is primarily driven by drag reduction.

\begin{figure}[h]
  \centering
  \begin{subfigure}[b]{0.49\textwidth}
    \centering
    \includegraphics[width=\linewidth]{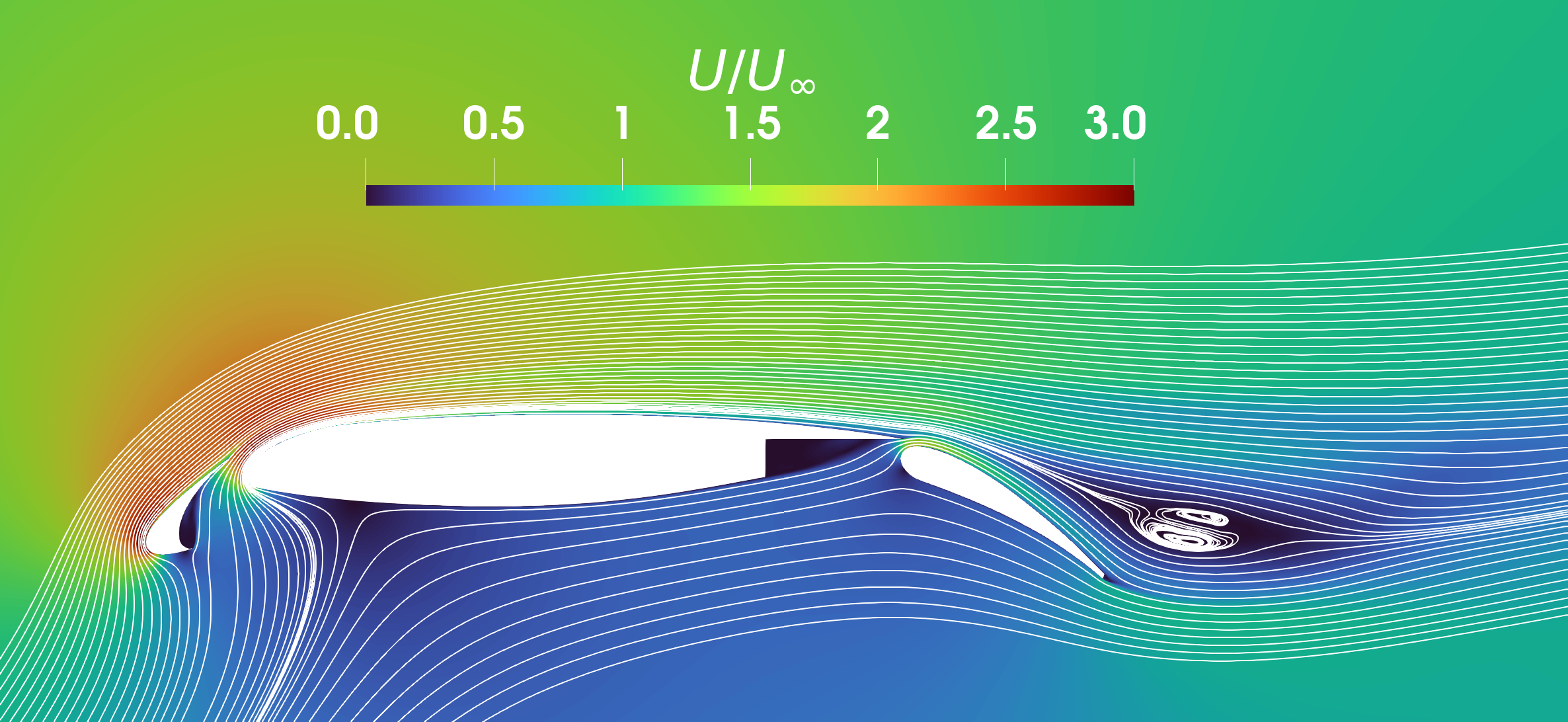}
    \caption{}
    \label{fig:BO_flow_a}
  \end{subfigure}
  \begin{subfigure}[b]{0.49\textwidth}
    \centering
    \includegraphics[width=\linewidth]{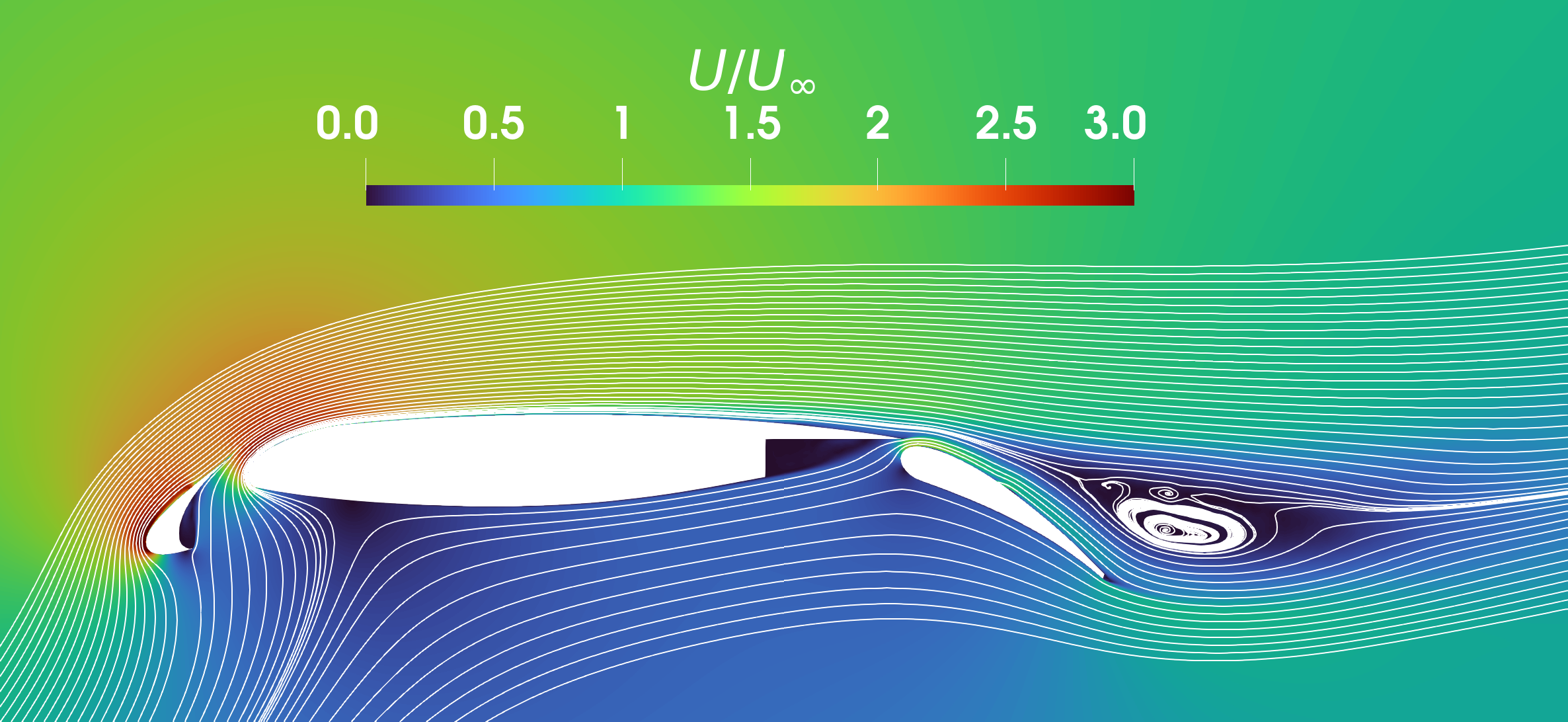}
    \caption{}
    \label{fig:BO_flow_b}
  \end{subfigure} 
  \caption{Streamlines of the time-averaged flow field with velocity magnitude in the background for the baseline scenario (left) and the BO-optimized AFC scenario (right) using the fine mesh.}
  \label{fig:BO_flow}
\end{figure}

Figure \ref{fig:BO_flow} shows the flow streamlines for the two analyzed cases, the non-actuated and BO-optimized AFC scenarios. The application of suction on the slat suction side reduces the boundary layer thickness, causing the slat wake to nearly disappear downstream. This is then a signature of drag reduction and hence, the slat element primarily drives the decrease in drag. However, the recirculation area above the flap is larger than in the baseline scenario, increasing the drag of the main and flap elements. Nevertheless, the net effect of the slat compensates for these increases, resulting in an overall drag reduction compared to the baseline scenario, and, consequently, an improvement in aerodynamic efficiency in the AFC configuration.

\subsection{DRL optimization} \label{sec:DRL_results}

In this section, the slat and main jet velocities are determined through an optimization process based on the training of a DRL agent, as described in Section \ref{sec:DRLsetup}. The objective is to demonstrate that a closed-loop control strategy, where actions are selected according to the instantaneous flow-field state, can outperform a conventional open-loop optimization with a fixed actuation law, such as the one presented in Section \ref{sec:BO_results}. This richer agent-environment interaction is expected to exploit more complex flow dynamics, potentially leading to greater improvements in aerodynamic performance, especially considering the MARL framework used, which enables three-dimensional actuations.

\begin{figure}[h]
  \centering
  \begin{subfigure}[b]{0.32\textwidth}
    \centering
    \includegraphics[width=\linewidth]{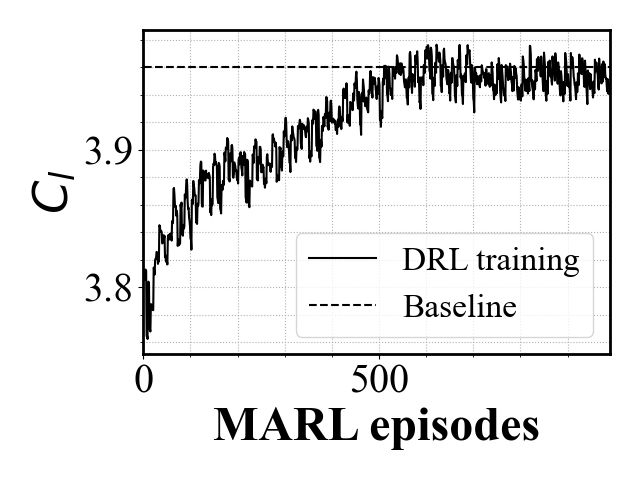}
    \caption{}
    \label{fig:DRL_train_a}
  \end{subfigure}
  \begin{subfigure}[b]{0.32\textwidth}
    \centering
    \includegraphics[width=\linewidth]{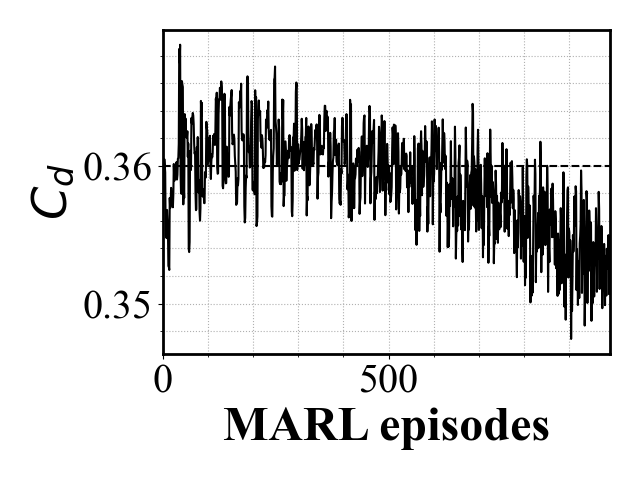}
    \caption{}
    \label{fig:DRL_train_b}
  \end{subfigure}
  \begin{subfigure}[b]{0.32\textwidth}
    \centering
    \includegraphics[width=\linewidth]{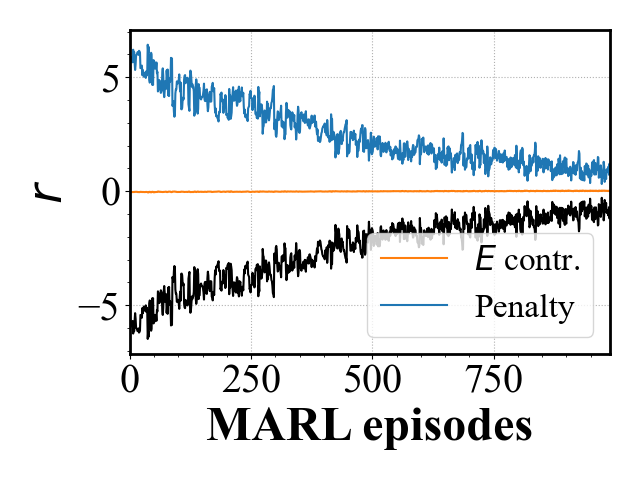}
    \caption{}
    \label{fig:DRL_train_c}
  \end{subfigure}
  \caption{Lift coefficient $C_l$ (left), drag coefficient $C_d$ (middle) and local reward $r$ (right) evolution during the DRL training. All quantities are averaged over the last $4 \, U_\infty/c$ of each episode and values from each MARL episode are concatenated sequentially.}
  \label{fig:DRL_train}
\end{figure}

Figure \ref{fig:DRL_train} illustrates the evolution of the lift coefficient $C_l$, drag coefficient $C_d$, and reward $r$ during the DRL training. In total, 990 MARL episodes are performed. This value considers all the environments $n_\mathrm{env} = 30$ executed in parallel as individual episodes, which translates into $990/30 = 33$ policy updates. As training progresses, $C_l$ increases until reaching a plateau in the final episodes where, although a few episodes slightly outperform the baseline case, the average $C_l$ remains below the baseline value. The drag coefficient $C_d$, which initially matches the baseline level, shows a gradual reduction toward the end of the training.

Consequently, the reward $r$ increases over the course of training, indicating a certain degree of learning. However, this improvement primarily stems from the reduction of penalty terms in the reward (Eqs. \ref{eq:target_penaltyCl} and \ref{eq:target_penaltyCd}) rather than from the improvement of aerodynamic efficiency $E$, which is the main goal of the selected reward. Specifically, as the drag constraint ($C_d < C_{d,\mathrm{base}}$) is satisfied at roughly the middle of the training, the subsequent reward growth is mainly due to a reduction in the lift penalty term. Nevertheless, since the lift coefficient asymptotically approaches but never surpasses the baseline value, the penalty term remains nonzero. As a result, no significant improvement in aerodynamic efficiency is achieved, which stays nearly constant throughout the training. The overall trend thus suggests that the penalty cannot be eliminated, preventing any increase in the efficiency term.

The hard constraint imposed within the reward function may limit the agent's ability to explore and progress toward regions of the optimization space that could enhance efficiency. This might create a local region from which the agent cannot escape once it converges there. A more continuous reward formulation might be a better selection for DRL agents and consequently, the reward design may not be the optimal in this case. As observed in other DRL studies, the final performance is highly sensitive to the reward formulation \cite{montala2025_drl_laminar}. Small modifications, such as adjusting the weighting factors of the penalty terms, can substantially alter the exploration landscape, leading to distinct solution manifolds and training outcomes.

Nevertheless, the practical computational limits of the current framework are being reached. While previous wing-based trainings required approximately 10 minutes per episode at $Re_c = 1{,}000$ \cite{montala2025_drl_laminar} and 120 minutes per episode at $Re_c = 60{,}000$ \cite{montala2025_drl_turb}, the present simulations require about 800 minutes per episode. This represents an exponential increase in wall-clock time with the Reynolds number. Considering the total number of 990 MARL episodes and the $n_\mathrm{env} =30$ parallel environments being used, the overall training duration is approximately two weeks. The increase in computational cost primarily arises from the smaller time steps required at higher Reynolds numbers. The training was conducted in 30 nodes of the MN5 supercomputer cluster located at the Barcelona Supercomputing Center, each equipped with four NVIDIA H100 GPUs and two Intel  XEON processors (80 CPU cores in total), yielding a total computational cost of approximately 48,000 GPU-h.

Given this computational expense, there is little room for hyperparameter tuning or testing alternative reward functions, and strategies to accelerate the training process must be considered. Several approaches can be envisioned. Even though the computational mesh cannot be further coarsened as it already lies within the acceptable resolution limits for LES, RANS-based training could be explored, as only time-averaged aerodynamic coefficients are required for learning.

Another possibility is to increase the number of domain partitions (to allocate more GPUs per CFD simulation) or to increase the number of parallel CFD simulations. Regarding the former, previous experience indicates that the minimum wall-clock time in SOD2D is achieved with approximately one million degrees of freedom (DOF) per GPU, whereas the present setup employs 2.5 million DOF per GPU. Concerning the latter, the DRL framework exhibits excellent scalability with the number of parallel CFD simulations. Although the upper limit of this scalability is still unknown, increasing from the current $n_\mathrm{CFD}=10$ to around 30 parallel CFD runs is expected to be feasible. These two strategies could yield an approximate $6\times$ reduction in wall-clock time while maintaining the overall computational cost.

Another potential strategy is the incorporation of surrogate models, which could replace high-fidelity CFD simulations in selected episodes. These surrogate-based episodes could be complemented by periodic high-fidelity simulations to refine both the surrogate model and the DRL agent simultaneously. Finally, transfer learning could also be employed by initially training at lower Reynolds numbers. This would significantly reduce the mesh size and time-step constraints, which otherwise increase exponentially with $Re$.

\section{Conclusions}
Wall-resolved large-eddy simulations (LES) of the 30P30N high-lift wing at $Re_c = 450{,}000$ and $\alpha=23^\circ$ were performed to evaluate active flow control (AFC) strategies using synthetic jets. Two optimization approaches were considered: Bayesian optimization (BO) for open-loop control and deep reinforcement learning (DRL) for closed-loop control, both aiming to mitigate stall and improve aerodynamic efficiency by adjusting the slat, main, and flap jet velocities.

The uncontrolled baseline was validated against literature data, showing good agreement in pressure distributions and aerodynamic coefficients. The fine mesh provided wall-resolved LES accuracy, while the coarser mesh offered reliable predictions at lower computational cost, enabling efficient optimization.

BO successfully identified an optimal steady jet configuration, increasing aerodynamic efficiency by +10.9\% relative to the baseline, maintaining lift, and reducing drag by -9.7\%. Flow analysis showed that suction on the slat element primarily drove the improvement by thinning the boundary layer and reducing wake intensity, while drag on the main and flap elements increased slightly. These results demonstrate BO's ability to efficiently explore the control space and identify interpretable near-optimal AFC strategies with few CFD evaluations.

In contrast, the DRL-based closed-loop control, despite using instantaneous flow information from distributed sensors, did not significantly improve efficiency. The agent slightly reduced drag and nearly maintained lift, but the reward was dominated by penalty terms, limiting gains in aerodynamic performance. This highlights the importance of reward design in DRL and suggests smoother formulations could enhance learning. However, high computational cost remains a bottleneck, restricting hyperparameter tuning and reward exploration. Future work should focus on accelerating training through surrogate modeling, transfer learning, or improved parallelization to allow more extensive exploration of the control space.

\section*{Acknowledgments}
This research was partially funded by the Ministerio de Ciencia e Innovación of Spain (PID2023-150408OB-C21/C22). Work has also been co-funded by the European High Performance Computing Joint Undertaking (EuroHPC JU) and Germany, Italy, Slovenia, Spain, Sweden, and France under grant agreement No 101092621; and under grant PCI2022-135085-2 funded by MICIU/AEI/10.13039/ 501100011033 and by the “European Union NextGenerationEU/PRTR”. Simulations were supported by the Red Española de Supercomputación (IM-2024-2-0004 and IM-2024-3-0005) and EuroHPC JU (EHPC-REG-2024R01-038), granting access to MN5 at the Barcelona Supercomputing Center. We also acknowledge AGAUR for supporting the LS/CFD (2021 SGR 00902) and TUAREG (2021 SGR 01051) research groups. R. M. thanks AGAUR for the FI-SDUR grant (2022 FISDU 00066), and R. V. acknowledges financial support from ERC grant no. 2021-CoG-101043998, DEEPCONTROL. Views and opinions expressed are however those of the author(s) only and do not necessarily reflect those of the European Union or the European Research Council.

\bibliographystyle{iopart-num}
\bibliography{references}

@article{Thibert1995,
  title={ONERA activities on high-lift devices fortransport aircraft},
  volume={99}, 
  DOI={https://doi.org/10.1017/S0001924000028700.},
  number={989}, 
  journal={The Aeronautical Journal}, 
  author={Thibert, J. J. and Reneaux, J. and Moens, F. and Preist, J.}, 
  year={1995}, 
  pages={395–411}
}

@techreport{Klausmeyer1997,
  author       = {Klausmeyer, Steven M. and Lin, John C.},
  title        = {Comparative Results from a {CFD} Challenge over a 2{D} Three-Element High-Lift Airfoil},
  institution  = {NASA Langley Research Center},
  number       = {NASA TM-112858},
  year         = {1997},
  month        = {May},
  note         = {{NASA} Technical Memorandum}
}

@inproceedings{Choudhari2015,
  author = {M. Choudhari and D. Lockard},
  title = {Assessment of Slat Noise Predictions for {30P30N} High-Lift Configuration from {BANC-III} Workshop},
  booktitle = {21st AIAA/CEAS Aeroacoustics Conference},
  year = {2015},
  doi = {https://doi.org/10.2514/6.2015-2844},
}

@article{Ashton2016,
  author = {N. Ashton and A. West and F. Mendon\c{c}a},
  title = "{Flow Dynamics Past a 30P30N Three-Element Airfoil Using Improved Delayed Detached-Eddy Simulation}",
  journal = {AIAA Journal},
  volume = {54},
  number = {11},
  pages = {3657-3667},
  year = {2016},
  doi = {https://doi.org/10.2514/1.J054521},
}

@inproceedings{Pascioni2014,
  author = {K. Pascioni and L. Cattafesta and M. Choudhari},
  title = {An Experimental Investigation of the {30P30N} Multi-Element High-Lift Airfoil},
  booktitle = {20th AIAA/CEAS Aeroacoustics Conference},
  year = {2014},
  doi = {https://doi.org/10.2514/6.2014-3062},
}

@article{Gao2020,
  author = {J. Gao and X. Li and D. Lin},
  title = "{Numerical Simulation of the 30P30N High-Lift Airfoil Noise with Spectral Difference Method}",
  journal = {AIAA Journal},
  volume = {58},
  number = {6},
  pages = {2517-2532},
  year = {2020},
  doi = {https://doi.org/10.2514/1.J058060},
}

@inproceedings{Ueno2019,
  author = {Y. Ueno and A. Ochi},
  title = {Airframe Noise Prediction Using {N}avier-{S}tokes Code with Cartesian and Boundary-fitted Layer Meshes},
  booktitle = {25th AIAA/CEAS Aeroacoustics Conference},
  year = {2019},
  doi = {https://doi.org/10.2514/6.2019-2553},
}

@article{Shur2023,
  author = {Mikhail Shur and Mikhail Strelets and Philippe Spalart and Andrey Travin},
  title = {Advanced detached-eddy simulation of the MD 30P-30N three-element airfoil},
  journal = {Journal of Turbulence},
  volume = {24},
  number = {11-12},
  pages = {554--576},
  year = {2023},
  publisher = {Taylor \& Francis},
  doi = {https://doi.org/10.1080/14685248.2023.2278506},
}

@article{Montala2024,
    author = {Montalà, R. and Lehmkuhl, O. and Rodriguez, I.},
    title = {On the dynamics of the turbulent flow past a three-element wing},
    journal = {Physics of Fluids},
    volume = {36},
    number = {2},
    pages = {025125},
    year = {2024},
    month = {02},
    issn = {1070-6631},
    doi = {https://doi.org/10.1063/5.0182215},
}

@article{Montala2025,
    author = {Montalà, R. and Lehmkuhl, O. and Rodriguez, I.},
    title = {On the Flow Past a Three-Element Wing: Mean Flow and Turbulent Statistics},
    journal = {Flow, Turbulence and Combustion},
    volume = {115},
    pages = {51-77},
    year = {2025},
    doi = {https://doi.org/10.1007/s10494-024-00566-y},
}

@Inbook{Piomelli1996,
  author={U. Piomelli and J. R. Chasnov},
  title="{Large-Eddy Simulations: Theory and Applications}",
  bookTitle="{Turbulence and Transition Modelling: Lecture Notes from the ERCOFTAC/IUTAM Summerschool held in Stockholm, 12--20 June, 1995}",
  year="1996",
  publisher="Springer Netherlands",
  address="Dordrecht",
  pages="269--336",
  isbn="978-94-015-8666-5",
  doi="https://doi.org/10.1007/978-94-015-8666-5_7",
}

@article{Vreman2004,
author = {A. W. Vreman},
title = "{An eddy-viscosity subgrid-scale model for turbulent shear flow:  Algebraic theory and applications}",
journal = {Physics of Fluids},
volume = {16},
number = {10},
pages = {3670-3681},
year = {2004},
doi = {https://doi.org/10.1063/1.1785131},
}

@article{Gasparino2024,
title = {{SOD2D}: A GPU-enabled Spectral Finite Elements Method for compressible scale-resolving simulations},
journal = {Computer Physics Communications},
volume = {297},
pages = {109067},
year = {2024},
issn = {0010-4655},
doi = {https://doi.org/10.1016/j.cpc.2023.109067},
author = {L. Gasparino and F. Spiga and O. Lehmkuhl},
}

@article{Shmilovich2009,
author = {Shmilovich, Arvin and Yadlin, Yoram},
title = {Active Flow Control for Practical High-Lift Systems},
journal = {Journal of Aircraft},
volume = {46},
number = {4},
pages = {1354-1364},
year = {2009},
doi = {https://doi.org/10.2514/1.41236},
}

@misc{Guadarrama2018,
  title = {{TF-Agents}: A library for reinforcement learning in {T}ensor{F}low},
  year = {2018},
  url = {https://github.com/tensorflow/agents},
  author = {S. Guadarrama and A. Korattikara and O. Ramirez and P. Castro and E. Holly and S. Fishman and K. Wang and E. Gonina and N. Wu and E. Kokiopoulou and L. Sbaiz and J. Smith and G. Bartók and J. Berent and C. Harris and V. Vanhoucke and E. Brevdo},
}

@article{Partee2022,
title = {Using machine learning at scale in numerical simulations with {S}mart{S}im: An application to ocean climate modeling},
journal = {J. Comput. Sci.},
volume = {62},
pages = {101707},
year = {2022},
issn = {1877-7503},
doi = {10.1016/j.jocs.2022.101707},
author = {S. Partee and M. Ellis and A. Rigazzi and A.E. Shao and S. Bachman and G. Marques and B. Robbins},
}

@article{Belus2019,
    author = {V. Belus and J. Rabault and J. Viquerat and Z. Che and E. Hachem and U. Reglade},
    title = {Exploiting locality and translational invariance to design effective deep reinforcement learning control of the 1-dimensional unstable falling liquid film},
    journal = {AIP Advances},
    volume = {9},
    number = {12},
    pages = {125014},
    year = {2019},
    month = {12},
    issn = {2158-3226},
    doi = {10.1063/1.5132378},
}

@article{Suarez2025a,
    author = {P. Suárez and F. Alcántara-Ávila and J. Rabault and A. Miró and B. Font and O. Lehmkuhl and R. Vinuesa},
    title = {Flow control of three-dimensional cylinders transitioning to turbulence via multi-agent reinforcement learning},
    journal = {Commun. Eng.},
    volume = {4},
    number = {113},
    year = {2025},
    doi = {10.1038/s44172-025-00446-x},
}

@article{Suarez2025b,
    author = {P. Suárez and F. Alcántara-Ávila and J. Rabault and A. Miró and B. Font and O. Lehmkuhl and R. Vinuesa},
    title = { Active Flow Control for Drag Reduction Through Multi-agent Reinforcement Learning on a Turbulent Cylinder at ${R}e_{D}=3900$},
    journal = {Flow Turbul. Combust.},
    volume = {115},
    pages={3--27},
    year = {2025},
    doi = {10.1007/s10494-025-00642-x},
}

@article{Font2025,
  title={Deep Reinforcement Learning for Active Flow Control in a Turbulent Separation Bubble},
  author={B. Font and F. Alc{\'a}ntara-{\'A}vila and J. Rabault and R. Vinuesa and O. Lehmkuhl},
  journal={Nat. Commun.},
  volume={16},
  number={1},
  pages={1422},
  year={2025},
  publisher={Nature Publishing Group UK London}
}

@misc{Schulman2017,
  author       = {J. Schulman and F. Wolski and P. Dhariwal and A. Radford and O. Klimov},
  title        = {Proximal policy optimization algorithms},
  year         = {2017},
  howpublished = {\url{https://doi.org/10.48550/arXiv.1707.06347}},
}

@article{You2008,
title = {Active control of flow separation over an airfoil using synthetic jets},
journal = {Journal of Fluids and Structures},
volume = {24},
number = {8},
pages = {1349-1357},
year = {2008},
note = {Unsteady Separated Flows and their Control},
issn = {0889-9746},
doi = {https://doi.org/10.1016/j.jfluidstructs.2008.06.017},
author = {D. You and P. Moin},
}

@article{Melton2006,
title = {Active control of separation from the flap of a supercritical airfoil},
journal = {AIAA Journal},
volume = {44},
number = {1},
pages = {012017},
year = {2006},
author = {Melton, L.P. and Yao, C.S. and Seifert, A.},
doi = {https://doi.org/10.2514/1.12225},
}

@article{Rodriguez2020,
title = { Effects of the Actuation on the Boundary Layer of an Airfoil at Reynolds Number Re = 60000},
journal = {Flow, Turbulence and Combustion},
volume = {105},
pages = {607-626},
year = {2020},
doi = {https://doi.org/10.1007/s10494-020-00160-y},
author = {I. Rodriguez and O. Lehmkuhl and R. Borrell},
}

@article{Lehmkuhl2020,
doi = {10.1088/1742-6596/1522/1/012017},
year = {2020},
month = {apr},
publisher = {IOP Publishing},
volume = {1522},
number = {1},
pages = {012017},
author = {Lehmkuhl, Oriol and Lozano-Durán, Adrián and Rodriguez, Ivette},
title = {Active flow control for external aerodynamics: from micro air vehicles to a full aircraft in stall},
journal = {Journal of Physics: Conference Series},
}

@article{Rabault2019,
  title={Artificial neural networks trained through deep reinforcement learning discover control strategies for active flow control},
  volume={865},
  DOI={10.1017/jfm.2019.62},
  journal={J. Fluid Mech.},
  author={J. Rabault and M. Kuchta and A. Jensen and U. Réglade and N. Cerardi},
  year={2019},
  pages={281--302}}

@article{Rabault_Kuhnle2019,
    author = {J. Rabault and A. Kuhnle},
    title = {Accelerating deep reinforcement learning strategies of flow control through a multi-environment approach},
    journal = {Phys. Fluids},
    volume = {31},
    number = {9},
    pages = {094105},
    year = {2019},
    month = {09},
    issn = {1070-6631},
    doi = {10.1063/1.5116415},
}

@article{Guastoni2023,
    author = {L. Guastoni and J. Rabault and P. Schlatter and H. Azizpour and R. Vinuesa},
    title = {Deep reinforcement learning for turbulent drag reduction in channel flows},
    journal = {Eur. Phys. J. E},
    volume = {46},
    pages={27},
    year = {2023},
    doi = {10.1140/epje/s10189-023-00285-8},
}

@article{Vasanth2024,
    author = {J. Vasanth and J. Rabault and F. Alcántara-Ávila and M. Mortensen and R. Vinuesa},
    title = {Multi-agent reinforcement learning for the control of three-dimensional {R}ayleigh–{B}énard convection},
    journal = {Flow Turbul. Combust.},
    year = {2024},
    doi = {10.1007/s10494-024-00619-2},
}

@article{Wang2022,
    author = {Y. Wang and Y. Mei and N. Aubry and Z. Chen and P. Wu and W. Wu},
    title = {Deep reinforcement learning based synthetic jet control on disturbed flow over airfoil},
    journal = {Phys. Fluids},
    volume = {34},
    number = {3},
    pages = {033606},
    year = {2022},
    month = {03},
    doi = {10.1063/5.0080922},
}

@article{Garcia2025,
title = {Deep-reinforcement-learning-based separation control in a two-dimensional airfoil},
journal = {Int. J. Heat Fluid Flow},
volume = {116},
pages = {109913},
year = {2025},
issn = {0142-727X},
doi = {10.1016/j.ijheatfluidflow.2025.109913},
author = {X. Garcia and A. Miró and P. Suárez and F. Alcántara-Ávila and J. Rabault and B. Font and O. Lehmkuhl and R. Vinuesa},
}

@misc{montala2025_drl_laminar,
  author = {Ricard Montalà and Bernat Font and Pau Suárez and Julien Rabault and Oliver Lehmkuhl and Roberto Vinuesa and Ignasi Rodriguez},
  title = {Deep Reinforcement Learning for Active Flow Control around a Three-Dimensional Flow-Separated Wing at {R}e = 1,000},
  year = {2025},
  howpublished = {\url{https://doi.org/10.48550/arXiv.2509.10195}},
}

@misc{montala2025_drl_turb,
  author = {Ricard Montalà and Bernat Font and Pau Suárez and Julien Rabault and Oliver Lehmkuhl and Roberto Vinuesa and Ignasi Rodriguez},
  title = {Discovering Flow Separation Control Strategies in 3{D} Wings via Deep Reinforcement Learning},
  year = {2025},
  howpublished = {\url{https://doi.org/10.48550/arXiv.2509.10185}},
}

@article{Han2023,
    author = {Han, Bing-Zheng and Huang, Wei-Xi and Xu, Chun-Xiao},
    title = {Multi-fidelity Bayesian optimization for spatially distributed control of flow over a circular cylinder},
    journal = {Physics of Fluids},
    volume = {35},
    number = {11},
    pages = {115144},
    year = {2023},
    month = {11},
    issn = {1070-6631},
    doi = {10.1063/5.0175403},
}

@article{Li2024,
  title={Jet mixing enhancement with Bayesian optimization, deep learning and persistent data topology},
  volume={991},
  DOI={10.1017/jfm.2024.525},
  journal={Journal of Fluid Mechanics},
  author={Li, Yiqing and Noack, Bernd R. and Wang, Tianyu and Cornejo Maceda, Guy Y. and Pickering, Ethan and Shaqarin, Tamir and Tyliszczak, Artur},
  year={2024},
  pages={A5}
}

@article{Morita2022,
title = {Applying Bayesian optimization with Gaussian process regression to computational fluid dynamics problems},
journal = {Journal of Computational Physics},
volume = {449},
pages = {110788},
year = {2022},
issn = {0021-9991},
doi = {https://doi.org/10.1016/j.jcp.2021.110788},
author = {Y. Morita and S. Rezaeiravesh and N. Tabatabaei and R. Vinuesa and K. Fukagata and P. Schlatter},
}

@article{Mahfoze2019,
  title = {Reducing the skin-friction drag of a turbulent boundary-layer flow with low-amplitude wall-normal blowing within a Bayesian optimization framework},
  author = {Mahfoze, O. A. and Moody, A. and Wynn, A. and Whalley, R. D. and Laizet, S.},
  journal = {Phys. Rev. Fluids},
  volume = {4},
  issue = {9},
  pages = {094601},
  numpages = {23},
  year = {2019},
  month = {Sep},
  publisher = {American Physical Society},
  doi = {10.1103/PhysRevFluids.4.094601},
}

\end{document}